\documentclass[prd,,aps,showpacs,nofootinbib,amsmath,amssymb]{revtex4}
\usepackage{amssymb,epsfig}
\usepackage{mathrsfs}
\usepackage[usenames,dvipsnames]{color}
\usepackage[bookmarks]{hyperref}

\newcommand{\Integer}{\mathbb{Z}}
\newcommand{\Real}{\mathbb{R}}

\newcommand{\dvol}{\mbox{dvol}}

\newcommand{\DD}{\mathbb{D}}
\newcommand{\EE}{\mathbb{E}}
\newcommand{\FF}{\mathbb{F}}
\newcommand{\HH}{\mathbb{H}}

\newcommand{\KK}{\mathbb{K}}

\newcommand{\proofof}[1]{\noindent {\bf Proof of #1. }}
\newcommand{\qed}{\hfill \fbox{} \vspace{.3cm}}

\newtheorem{theorem}{Theorem}

\begin{document}

\title{Phase space mixing in the equatorial plane of a Kerr black hole}

\author{Paola Rioseco and Olivier Sarbach}
\affiliation{$^1$Instituto de F\'isica y Matem\'aticas,
Universidad Michoacana de San Nicol\'as de Hidalgo,
Edificio C-3, Ciudad Universitaria, 58040 Morelia, Michoac\'an, M\'exico.}

\begin{abstract}
It is shown that a collisionless, relativistic kinetic gas configuration propagating in the equatorial plane of a Kerr black hole undergoes a relaxation process and eventually settles down to a stationary, axisymmetric configuration surrounding the black hole. The underlying mechanism for this relaxation process is due to phase space mixing, which implies that although the one-particle distribution function $f$ satisfying the collisionless Boltzmann equation is quasi-periodic in time, the associated macroscopic observables computed from averages over $f$ possess well-defined limits as time goes to infinity. The final state of the gas is described by an effective distribution function depending only upon the constants of motion, and it can be determined by an appropriate average of the initial distribution function.
\end{abstract}

\date{\today}


\maketitle

\section{Introduction}

Phase space mixing plays an important role in a wide range of areas in physics, including galactic dynamics, plasma physics and quantum physics. Roughly speaking, this phenomenon can be understood as the relaxation of the observables associated with a distribution function which is transported along an anharmonic Hamiltonian flow and hence spreads all over phase space due to different orbits having different angular frequencies.

In the context of galactic dynamics, groundbreaking work by Lynden-Bell~\cite{dL62,dL67} has shown that phase space mixing might be one of the fundamental mechanisms responsible for driving the one-particle distribution function describing the stellar distribution in a galaxy to an equilibrium configuration, although collisions or the exchange of energy between individual stars are negligible. For further important discussions on these topics, see for instance Refs.~\cite{sTmHdL86,sT99,BinneyTremaine-Book,gCrSmFbGpKpA14}.

In plasma physics, phase mixing has been argued to be the driving mechanism for Landau damping, the relaxation of a charged collisionless gas to a homogeneous configuration. Recently, this explanation has been put on a rigorous basis by the work of Mouhot and Villani~\cite{cMcV11} who established that for the case of a finite box with periodic boundary conditions the known phase-mixing property for the linearized Vlasov-Poisson also occurs in the full Vlasov-Poisson system, without linearization. For generalization of this work to the special-relativistic setting, see~\cite{bY15,bY16}.

At the quantum level, phase space mixing has recently been applied~\cite{rMeT17} to quenches in Bose-Einstein condensates in order to understand the behavior of the condensate in the vicinity of the saddle point in a double-well potential. Again, it was found that the system relaxes to a steady state due to phase-space mixing. For field-theoretical applications of the mixing phenomenon, see for example~\cite{tDaKeKyS02,tDaKnRyS02}.

In the present work, we analyze the effects of phase mixing and its associated relaxation process in a general relativistic scenario. More specifically, we consider a collisionless, relativistic gas configuration that is trapped by the gravitational field of a rotating black hole. We restrict ourselves to the simplest case in which the gas configuration is sufficiently thin such that its self-gravity can be neglected and in which the gas is confined to the equatorial plane of the black hole, leaving the discussion of more realistic configurations to future work~\cite{pRoS18c}. As a consequence of our assumptions, each individual gas particle follows a bound geodesic trajectory in the equatorial plane of a Kerr black hole background, and an explicit solution representation for the one-particle distribution function can be obtained by representing the geodesic flow in terms of action-angle-like variables, see Section~\ref{Sec:KerrGeodesics}.

Based on this solution representation, we compute the particle current density four-vector and provide some examples in Section~\ref{Sec:Relaxation} showing that the particle density measured by a stationary observer outside the event horizon, although fluctuating in time, undergoes damped oscillations and eventually settles down to a constant value. An intuitive explanation for this convergence is given by exhibiting snapshots of the distribution function in the momentum space of the observer at different times which clearly illustrate the mixing in phase space.

To provide a precise mathematical formulation for the mixing property, in Section~\ref{Sec:Proof} we consider an observable $N[\varphi]$ which is obtained by integrating the one-particle distribution function over a given test function $\varphi$ on relativistic phase space. The time evolution of this observable is obtained by transporting the test function along the vector field generating the time translation symmetry. Denoting this transported test function by $\varphi_t$, we formulate and prove a theorem which shows that, provided a certain determinant condition holds on the support of $\varphi$, $N[\varphi_t]$ converges for $t\to \infty$ to the same observable $N[\varphi]$ with the distribution function replaced by its average over the angle variables. Therefore, apart from providing a rigorous formulation for the mixing property, our theorem allows to predict the final state of the gas configuration by considering the average of the initial distribution function. Finally, we show that the determinant condition is satisfied almost everywhere in phase space, and discuss the applicability of our theorem to the examples in Section~\ref{Sec:Relaxation}.

In Section~\ref{Sec:Conclusions} we present the conclusions and main implications of our results, and also give an outlook to future work. Technical details and relevant analytic expressions required for this article are listed in an appendix. Throughout this work we use geometrized units in which the gravitational constant and the speed of light are one.

\section{Collisionless distribution function in terms of action-angle variables}
\label{Sec:KerrGeodesics}

The geodesic flow describing the motion of free falling, massive particles following (spatially) bound trajectories on a Kerr spacetime may be represented analytically in terms of action-angle-like variables~\cite{wS02,tHeF08,rFwH09}. For the purpose of this work, it is sufficient to consider the $3$-dimensional spacetime $({\cal M},g)$ describing the induced geometry on the equatorial plane of a Kerr black hole exterior of mass $M > 0$ and rotational parameter $a$ satisfying $|a|\leq M$. In terms of Boyer-Lindquist coordinates $(t,r,\varphi)$, the metric has the following representation:
\begin{equation}
g = -dt^2 + \frac{2M}{r}\left( dt - a d\varphi \right)^2 + (r^2 + a^2) d\varphi^2 + \frac{r^2}{\Delta} dr^2,
\qquad r > r_H
\label{Eq:KerrBL}
\end{equation}
with $\Delta := r^2 - 2Mr + a^2$  and $r_H := M + \sqrt{M^2 - a^2}$ the radius of the event horizon. Due to the time-translation and axial symmetry of $({\cal M},g)$, the geodesic equations form an integrable Hamiltonian system which is characterized by the free-particle Hamiltonian ${\cal H}$ and the constants of motion ${\cal E}$ and ${\cal L}$, given by the following functions on the co-tangent  bundle $T^*{\cal M}$ associated with ${\cal M}$:
$$
{\cal H}(x,p) := \frac{1}{2} g^{\mu\nu}(x) p_\mu p_\nu,\quad
{\cal E}(x,p) := -p_t,\quad
{\cal L}(x,p) := p_\varphi,\qquad (x,p)\in T^*{\cal M},
$$
which Poisson-commute among themselves. The orbits are confined to the invariant subsets 
$$
\Gamma_{m,E,L} := \{ (x,p)\in T^*{\cal M} : {\cal H}(x,p) = -\frac{m^2}{2}, {\cal E}(x,p) = E, {\cal L}(x,p) = L \},
$$
with $m > 0$. Since only bound orbits are considered, the angular momentum $L$ has to be large enough in magnitude such that $L^2 > L_{ms}^2$, with $L_{ms}$ the angular momentum corresponding to the marginally stable circular orbit~\cite{jBwPsT72}, and the energy $E$ has to lie inside a certain interval $E_{min}(L) < E < E_{max}(L)$ with $E_{min}(L)$ the energy of the stable circular orbit with angular momentum $L$ and $E_{max}(L)\leq m$ the maximum of the potential well. In this range, it can be verified that the invariant sets $\Gamma_{m,E,L}$ are smooth $3$-dimensional submanifolds of $T^*{\cal M}$ having topology $\Real\times S^1\times S^1$. For the following, we focus on the phase space of bounded trajectories $\Gamma_{bound}$, the union of all these invariant submanifolds. 

Using standard tools from Hamiltonian mechanics~\cite{Arnold-Book} one can introduce action-angle variables $(Q^\alpha,J_\alpha)$ on $\Gamma_{bound}$, see for instance~\cite{tHeF08,wS02}. The action variables are defined as\footnote{Our definition of $J_0$ differs from the definition of $J_t$ in~\cite{tHeF08} by a minus sign.}
\begin{eqnarray}
J_0 := -\frac{1}{T}\int\limits_0^T p_t dt = {\cal E},\qquad
J_1 := \frac{1}{2\pi}\oint p_\varphi d\varphi = {\cal L},\qquad
J_2 := \frac{1}{2\pi}\oint p_r dr = \frac{1}{\pi}\int\limits_{r_1}^{r_2} \frac{\sqrt{R(r)}}{\Delta} dr,
\end{eqnarray}
where here the first integral defining $J_0$ is performed along an integral curve of the vector field $\partial_t$ ($t$ being the parameter along this curve), the second integral defining $J_1$ is similarly performed over the closed curve along the vector field $\partial_\varphi$,\footnote{More precisely, these integral curves are defined with respect to the complete lifts of the Killing vector fields $\partial_t$ and $\partial_\varphi$ on $T^*{\cal M}$, see for example Refs.~\cite{oStZ14b,pRoS16} for the most important properties of the complete lift in the context of relativistic kinetic theory.} and the third integral over the closed curve in the $(r,p_r)$-plane described by the radial equation $(\Delta p_r)^2 = R(r)$ with
\begin{equation}
R(r) := (E r^2 - a\hat{L})^2 - \Delta(m^2 r^2 + \hat{L}^2),\qquad
\hat{L} := L - a E,
\label{Eq:EnergySurface}
\end{equation}
and with $r_1 < r_2$ the turning points. The angle variables $Q^\alpha := \frac{\partial {\cal S}}{\partial J_\alpha}$ are obtained from the generating function
\begin{equation}
{\cal S}(x; J_0,J_1,J_2) = - E t + L\varphi + \int\limits_{(r,p_r)} p_r dr,
\label{Eq:GeneratingFunctionBis}
\end{equation}
where the integral on the right-hand side should be interpreted as a line integral along the curve $(\Delta p_r)^2 = R(r)$ connecting the reference point $(r_1,0)$ to the given point $(r,p_r)$ on this curve. ${\cal S}$, $Q^\alpha$, and $J_\alpha$ admit explicit representations in terms of standard elliptic integrals (details of the derivation will be presented elsewhere~\cite{pRoS18c}). In order to describe the result we denote the roots of the fourth-order polynomial $R(r)$ by $0 < r_0 < r_1 < r_2$ and introduce the dimensionless variables
$$
\alpha := \frac{a}{M},\qquad
\varepsilon := \frac{E}{m},\qquad
\lambda := \frac{L}{M m},\qquad
\xi_j := \frac{r_j}{M},\qquad j=0,1,2.
$$

\begin{figure}[ht]
\centerline{
\resizebox{10.6cm}{!}{\includegraphics{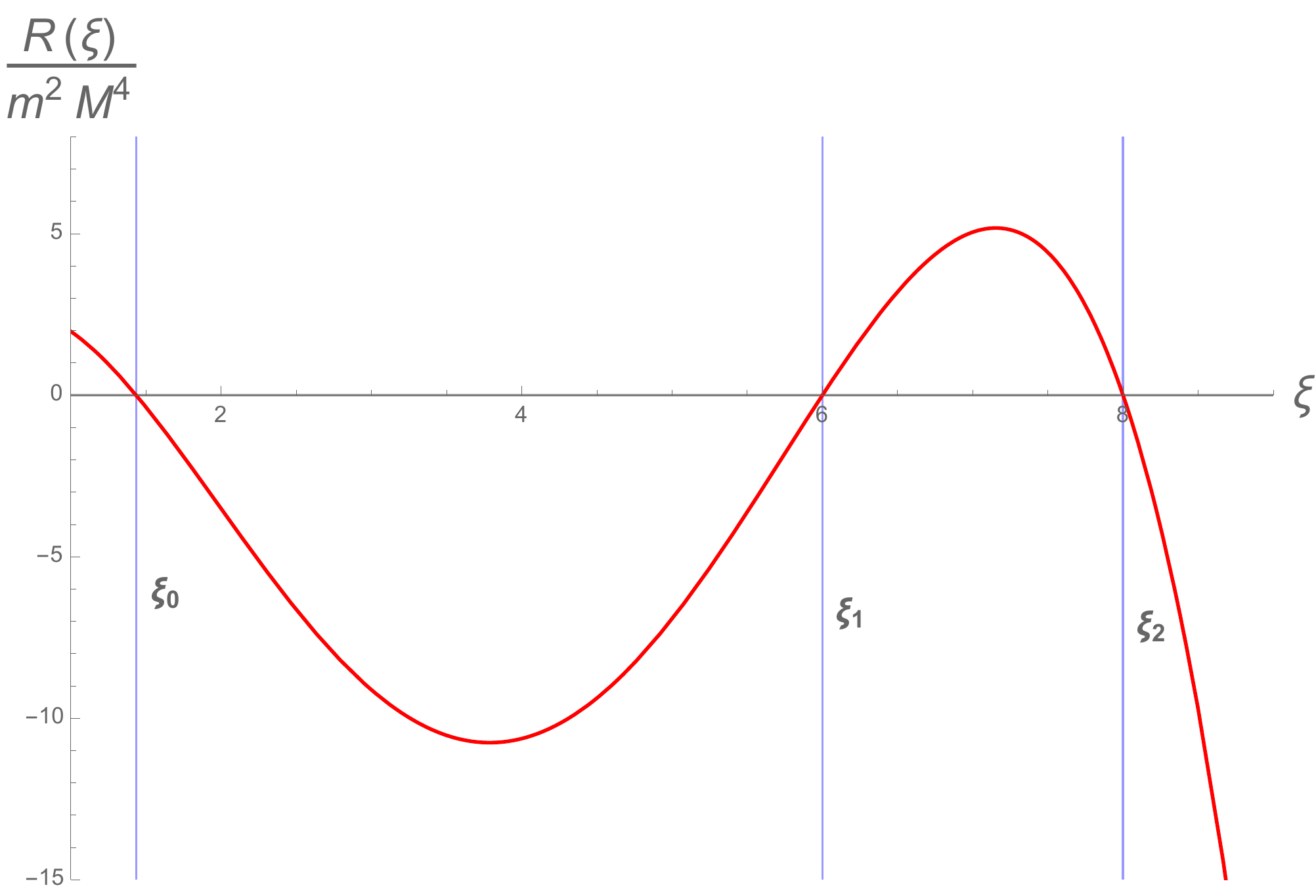}}
\resizebox{3.7cm}{!}{\includegraphics{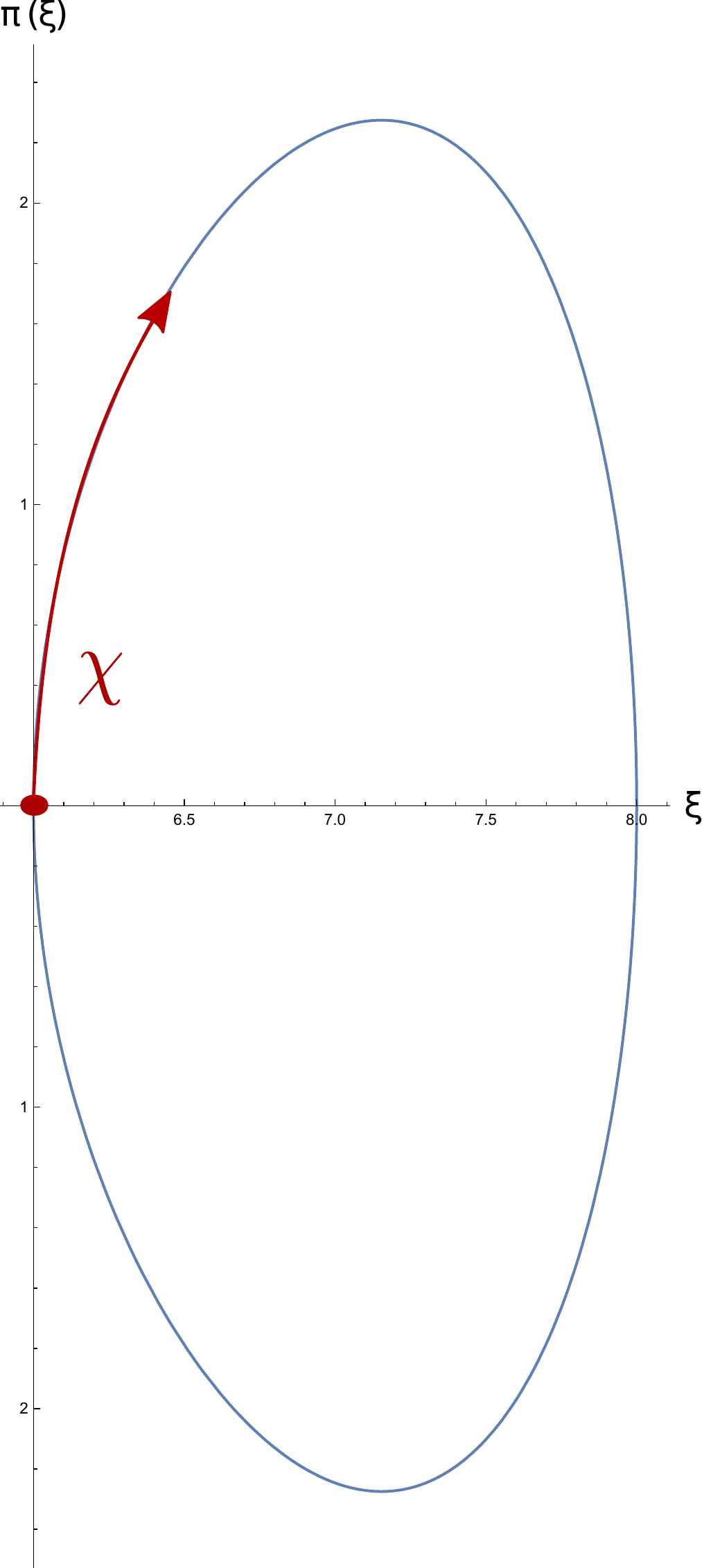}}
}
\caption{Left panel: The function $R(r)/M^4 m^2$ and its three non-trivial roots $\xi_0 < \xi_1 < \xi_2$ for the parameter values $\alpha = 0.9$, $\varepsilon = 0.93297$ and $\lambda = 2.9528$. Right panel: the projection of the set $\Gamma_{m,E,L}$ with the same parameter values onto the $(r,p_r)$-plane and the angle $\chi$ parametrizing this set. Here, $\xi := r/M$ and $\pi(\xi) := \Delta p_r/M^2 m$.
}
\label{Fig:parametric}
\end{figure}

The curve $(\Delta p_r)^2 = R(r)$ is parametrized by the $\pi$-periodic angle coordinate $\chi$ (see Fig.~\ref{Fig:parametric}) defined by
\begin{eqnarray}
\frac{r}{M} &=& \xi_0 + \frac{\xi_1 - \xi_0}{1 - b^2\sin^2\chi},\\
\frac{\Delta p_r}{M^2 m} 
 &=& \frac{1}{2}\sqrt{1 - \varepsilon^2}(\xi_2 - \xi_1)(\xi_1 - \xi_0)\sqrt{\frac{\xi_1}{\xi_2 - \xi_0}}
 \frac{\sqrt{1 - k^2\sin^2\chi}}{(1 - b^2\sin^2\chi)^2}\sin(2\chi),
\end{eqnarray}
with $b := \sqrt{(\xi_2 - \xi_1)/(\xi_2 - \xi_0)}$ and $k := \sqrt{\xi_0/\xi_1} b$, such that $0 < k < b < 1$. In terms of these quantities one finds
\begin{eqnarray}
&& J_0 = m\varepsilon,\qquad 
 J_1 = M m\lambda,\qquad 
 J_2 = \frac{M m}{\pi}\left[ (1-\varepsilon^2)\HH_0 + \varepsilon \HH_3 \right],
\label{Eq:JDef}\\
&& Q^0 = -t + M\frac{\HH_0\HH_2(\chi) - \HH_2\HH_0(\chi)}{\HH_0},\qquad
Q^1 = \varphi + \frac{\HH_0\HH_1(\chi) - \HH_1\HH_0(\chi)}{\HH_0},\qquad
Q^2 = \pi\frac{\HH_0(\chi)}{\HH_0},
\label{Eq:QDef}
\end{eqnarray}
where the functions $\HH_j(\chi)$ and corresponding constants $\HH_j := \HH_j(\pi/2)$ are defined in Appendix~\ref{App:HHFunctions}. The coordinates $(Q^\alpha,J_\alpha)$ provide new symplectic coordinates on $\Gamma_{bound}$, the submanifold of $T^*{\cal M}$ corresponding to bound orbits. The action variables $J_\alpha$ label the invariant submanifolds $\Gamma_{m,E,L}$, the coordinates $(Q^1,Q^2)$ are $2\pi$-periodic functions providing angles on each $S^1$-factor of $\Gamma_{m,E,L} = \Real\times S^1\times S^1$, while $Q^0$ parametrizes its $\Real$-factor.

In terms of these action-angle-like variables, the collisionless Boltzmann equation $\{ {\cal H}, f \} = 0$ for the one-particle distribution function $f$ assumes the simple form
\begin{equation}
\left( \Omega^0\frac{\partial}{\partial Q^0} + \Omega^1\frac{\partial}{\partial Q^1}
 + \Omega^2\frac{\partial}{\partial Q^2} \right) f = 0,\qquad
\Omega^\alpha := \frac{\partial {\cal H}}{\partial J_\alpha}.
\label{Eq:Boltzmann}
\end{equation}
Since ${\cal H}$ and $\Omega^\alpha$ only depend on the action variables $J_\alpha$, the most general solution of Eq.~(\ref{Eq:Boltzmann}) has the form
\begin{equation}
f(x,p) = F(Q^1 - \omega^1 Q^0, Q^2 - \omega^2 Q^0, J_0,J_1,J_2),
\label{Eq:GeneralSolution}
\end{equation}
with $F(Q^1,Q^2,J_0,J_1,J_2)$ an arbitrary function which is $2\pi$-periodic in the $Q$ variables and decays sufficiently fast in the $J$ variables, such that the integrals defining the spacetime observables are well-defined. Here the frequencies $\omega^1$ and $\omega^2$ are defined by
\begin{equation}
\omega^1 := \frac{\Omega^1}{\Omega^0} 
 = \frac{1}{M}\frac{\HH_1}{\HH_2 - \varepsilon\HH_0},\qquad
\omega^2 := \frac{\Omega^2}{\Omega^0} = -\frac{\pi}{\HH_1}\omega^1.
\label{Eq:FundamentalFrequencies}
\end{equation}
Eqs.~(\ref{Eq:JDef},\ref{Eq:QDef},\ref{Eq:GeneralSolution}) provide an explicit solution representation for the distribution function $f$ which will enable us to study the dynamical behavior of generic, time-dependent gas configurations in the next two sections. Note that according to Eq.~(\ref{Eq:QDef}) the distribution function $f$ is axisymmetric if and only if $F$ is independent of $Q^1$, and it is stationary and axisymmetric if and only if $F$ is independent of $Q^1$ and $Q^2$ (in which case it depends only on the action variables $J_\alpha$). In the remainder of this work, we demonstrate that any distribution function relaxes in time (in some sense made precise in Section~\ref{Sec:Proof}) to such a stationary and axisymmetric configuration.

\section{Relaxation of spacetime observables and phase space mixing}
\label{Sec:Relaxation}

In this section we analyze the behavior of spacetime observables along the world lines of observers located outside the event horizon. For simplicity and the sake of illustration, in this section we assume that the black hole is non-rotating (although we will come back to the rotating case in the next section). Furthermore, we focus on the particle current density four-vector (see for instance~\cite{jE71})
$$
{\cal J}^\mu(x) = \int f(x,p) p^\mu \dvol_x(p),\qquad
\dvol_x(p) := \sqrt{-\det(g^{\mu\nu}(x))} dp_t dp_r dp_\vartheta dp_\varphi,
$$
measured by a static observer in the equatorial plane with fixed spatial coordinates $(r,\vartheta,\varphi) = (r_{obs},\pi/2,\varphi_{obs})$, $r_{obs} > 2M$, with respect to its rest frame
$$
e_0 = \frac{1}{\sqrt{1-\frac{2M}{r}}}\frac{\partial}{\partial t},\qquad
e_1 = \sqrt{1-\frac{2M}{r}}\frac{\partial}{\partial r},\qquad
e_2 = \frac{1}{r}\frac{\partial}{\partial \vartheta},\qquad
e_3 = \frac{1}{r} \frac{\partial}{\partial \varphi}.
$$
In order to perform the integral over the momentum we first re-express the volume form $\dvol_x(p)$ in terms of the conserved quantities $m,E,\ell$ and $\ell_z$, where $\ell$ and $\ell_z$ are, respectively, the total and the azimuthal component of the angular momentum. For points $x$ located in the equatorial plane, one obtains
$$
\dvol_x(p) = \frac{1}{r^2} dE dp_r dp_{\vartheta} d \ell _z 
 = \frac{dE (m dm) (\ell d\ell) d\sigma}{\sqrt{R(r)}},
$$
where we have defined $\sigma$ by $\sin\sigma = \ell_z/\ell$ and the function $R(r)$ is given in Eq.~(\ref{Eq:EnergySurface}). Assuming a kinetic gas distribution of identical particles of positive rest mass $m$ which are confined to the equatorial plane (such that $\sigma = \pm \pi/2$ and $\ell_z = L$), one obtains the following orthonormal components of the current density:
\begin{equation}
{\cal J}^\alpha := \left. m^3\int 
\frac{ \left[ f(x,p_+) p_+^\alpha + f(x,p_-) p_-^\alpha\right] |\lambda| d\varepsilon d\lambda}
{\sqrt{(1-\varepsilon^2)\xi(\xi - \xi_0)(\xi - \xi_1)(\xi_2 - \xi)}} \right|_{\xi = r_{obs}/M},
\label{Eq:J0}
\end{equation}
with $p^0_\pm := m\varepsilon/\sqrt{1 - \frac{2}{\xi}}$, $p^1_\pm := \sqrt{1 - \frac{2}{\xi}} p_{r\pm}$, $p^2_\pm := 0$ and $p^3_\pm := m\lambda/\xi$, and where $p_\pm$ denote the two possible values for the four-momentum $p = p^\alpha e_\alpha$ corresponding to the two solutions of equation $(\Delta p_r)^2 = R(r)$. To make further progress, we use the relations~(\ref{relations_rootsE}) to re-express the integral in terms of the turning points $\xi_1$ and $\xi_2$. Instead of $\xi_1$ and $\xi_2$, it is convenient to use the generalized Keplerian variables $(P,e)$ as in Ref.~\cite{jBmGtH15}, which are defined by
$$
\xi_1 = \frac{P}{1 + e},\qquad
\xi_2 = \frac{P}{1 - e},
$$
such that $\xi_0 = 2P/(P-4)$. Here, the eccentricity $e$ is restricted to the interval $0 < e < 1$, the limit $e\to 0$ representing circular trajectories, and the semi-latus rectum $P$ is restricted to $P > 6 + 2e$, the limits $P\to \infty$ and $P\to 6 + 2e$ (with fixed $e$) corresponding, respectively, to the Newtonian limit and the innermost stable orbits (ISO) which separate the bound orbits from those that plunge into the black hole. Note that there is a one-to-two correspondence between the parameters $(P,e)$ and the constants of motion $(\varepsilon,\pm\lambda)$. Note also that given $\xi_{obs} = r_{obs}/M$, the parameters $(P,e)$ are restricted by the conditions $\xi_0 < \xi_1 < \xi_{obs} < \xi_2$, which yield
$$
P_{min}(e) := \max\{ 6 + 2e, (1-e)\xi_{obs} \} < P < (1+e)\xi_{obs} =: P_{max}(e).
$$
Based on these observations, Eq.~(\ref{Eq:J0}) can be rewritten as the sum ${\cal J}^\alpha = {\cal J}_{\lambda > 0}^\alpha + {\cal J}_{\lambda <  0}^\alpha$ over the contributions corresponding to gas particles with positive/negative angular momentum $\lambda$, where
\begin{equation}
{\cal J}_{\lambda > 0}^\alpha = \left. \frac{m^3}{2}\sum\limits_{\pm}
\int\limits_0^1 de \int\limits_{P_{min}(e)}^{P_{max}(e)} dP
\frac{ F(Q_\pm^1 - \omega^1 Q_\pm^0,Q_\pm^2 - \omega^2 Q_\pm^0,J_0,J_1,J_2) p^\alpha_\pm}
{\sqrt{(1-\varepsilon^2)\xi(\xi - \xi_0)(\xi - \xi_1)(\xi_2 - \xi)}} 
\frac{e\sqrt{P}[ (P-6)^2 - 4e^2]}{\sqrt{(P-3-e^2)^5[(P-2)^2 - 4e^2]}} \right|_{\xi = \xi_{obs}}.
\label{Eq:Jlambda}
\end{equation}
Here, it is understood that all the relevant quantities are expressed in terms of $(P,e)$ which determine the locations of the roots $\xi_0,\xi_1,\xi_2$ and the conserved quantities $(\varepsilon,\lambda)$ taking the positive sign of $\lambda$ (and similarly for ${\cal J}_{\lambda < 0}^0$ where one takes the negative sign of $\lambda$). The sum over the $\pm$ signs refers to the two possible choices for $p_\pm$ which, in turn correspond to the two possible values for the angle variables $Q_\pm^\alpha$. Note that $Q_+^\alpha + Q_-^\alpha = 2\pi$.

\begin{figure}[ht]
\centerline{
\resizebox{9.0cm}{!}{\includegraphics{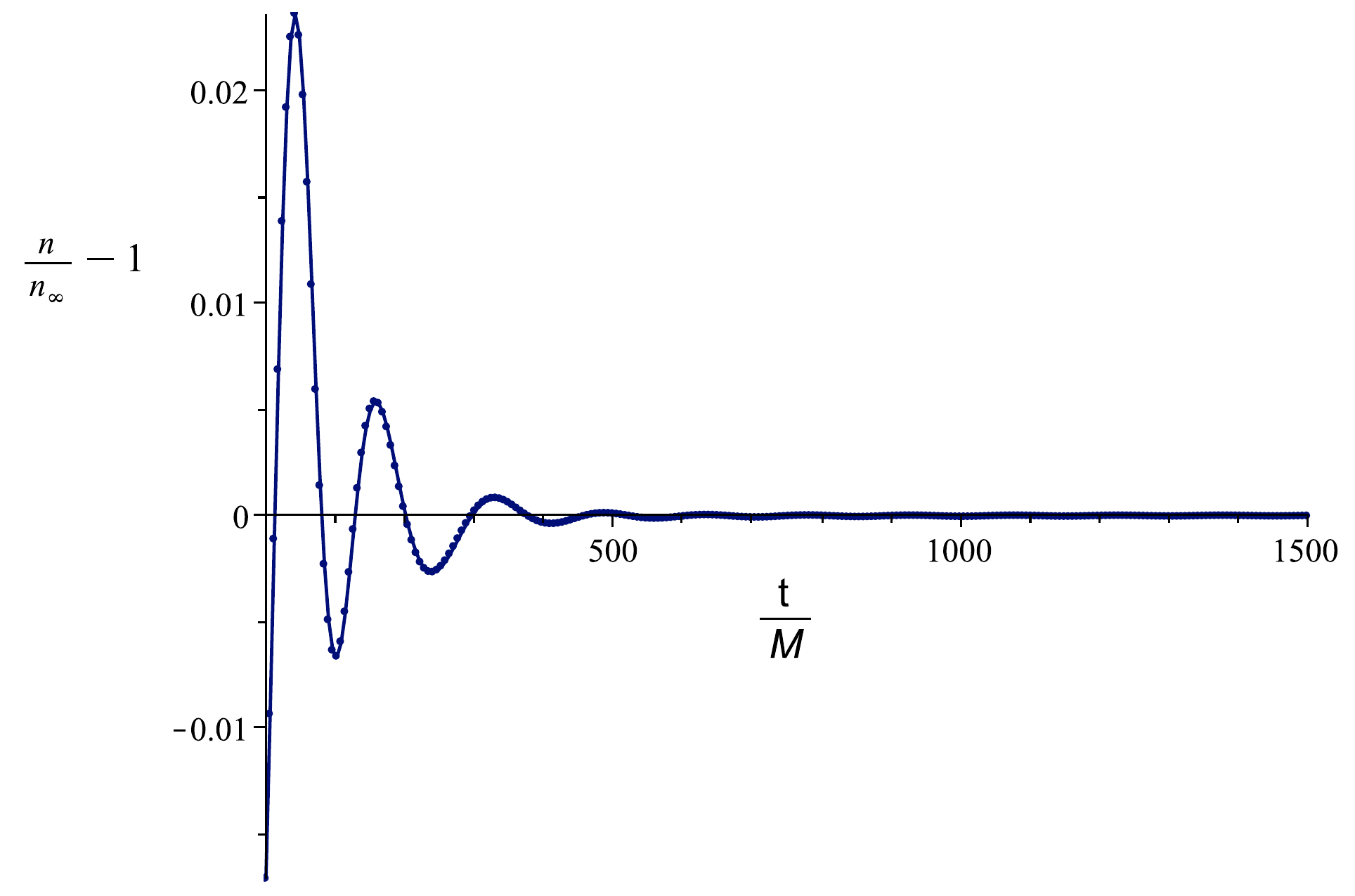}}
\resizebox{9.0cm}{!}{\includegraphics{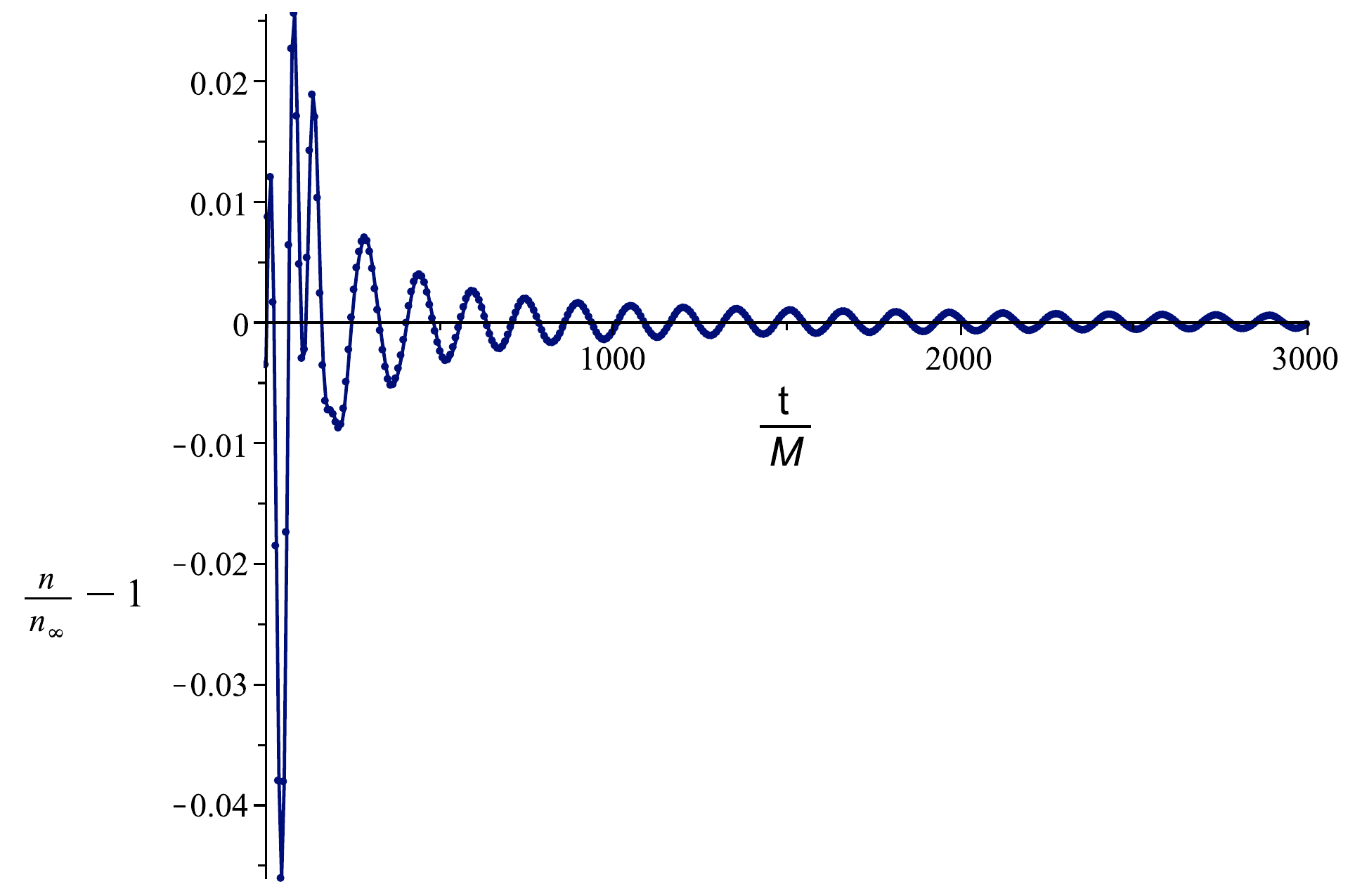}}
}
\caption{Particle density as a function of time measured by a static observer located at $r_{obs} = 6M$ and $\varphi_{obs} = 0$ in a Schwarzschild spacetime for case A (left panel) and case B (right panel). As is visible from these plots, the initial decay is much faster in case A, although the $\log-\log$ plots below indicate that in both cases the final decay is of the inverse power-law type.}
\label{Fig:nA}
\end{figure}

\begin{figure}[ht]
\centerline{
\resizebox{9.0cm}{!}{\includegraphics[trim= 36 180 50 60, clip=true]{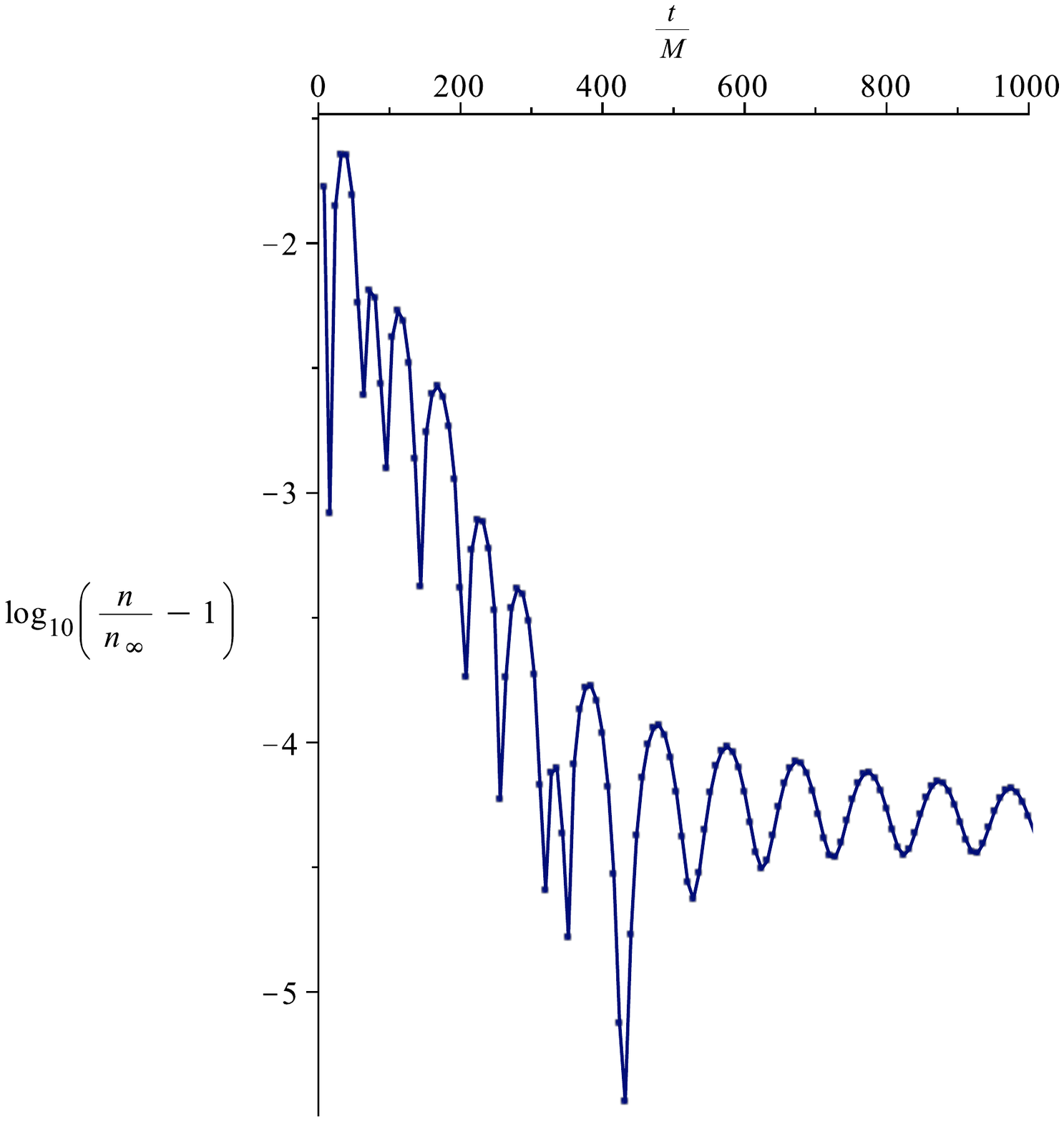}}
\resizebox{9.0cm}{!}{\includegraphics[trim= 36 180 50 60, clip=true]{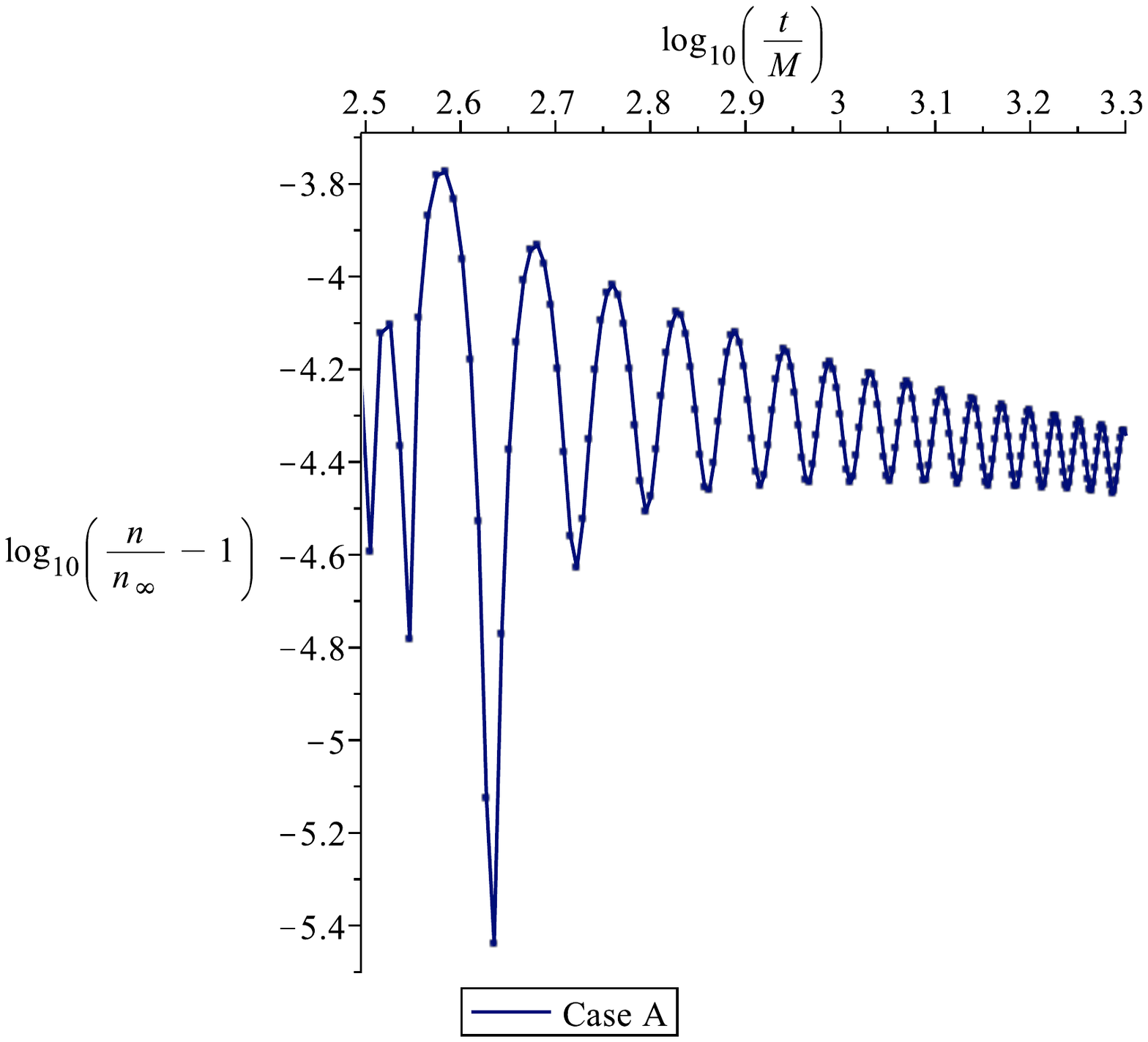}}
}
\caption{This plot shows the decay of the particle density towards its asymptotic value $n_\infty$ for case A. Left panel: $\log$ plot of the relative error $|n/n_\infty - 1|$, indicating an initial exponential decay for the initial period until $t\simeq 400M$. Right panel: $\log-\log$ plot of the relative error, indicating that for times larger than $\simeq 700M$ the decay is slower (apparently of the inverse power-law type).}
\label{Fig:nB}
\end{figure}

\begin{figure}[ht]
\centerline{
\resizebox{9.0cm}{!}{\includegraphics[trim= 36 160 36 60, clip=true]{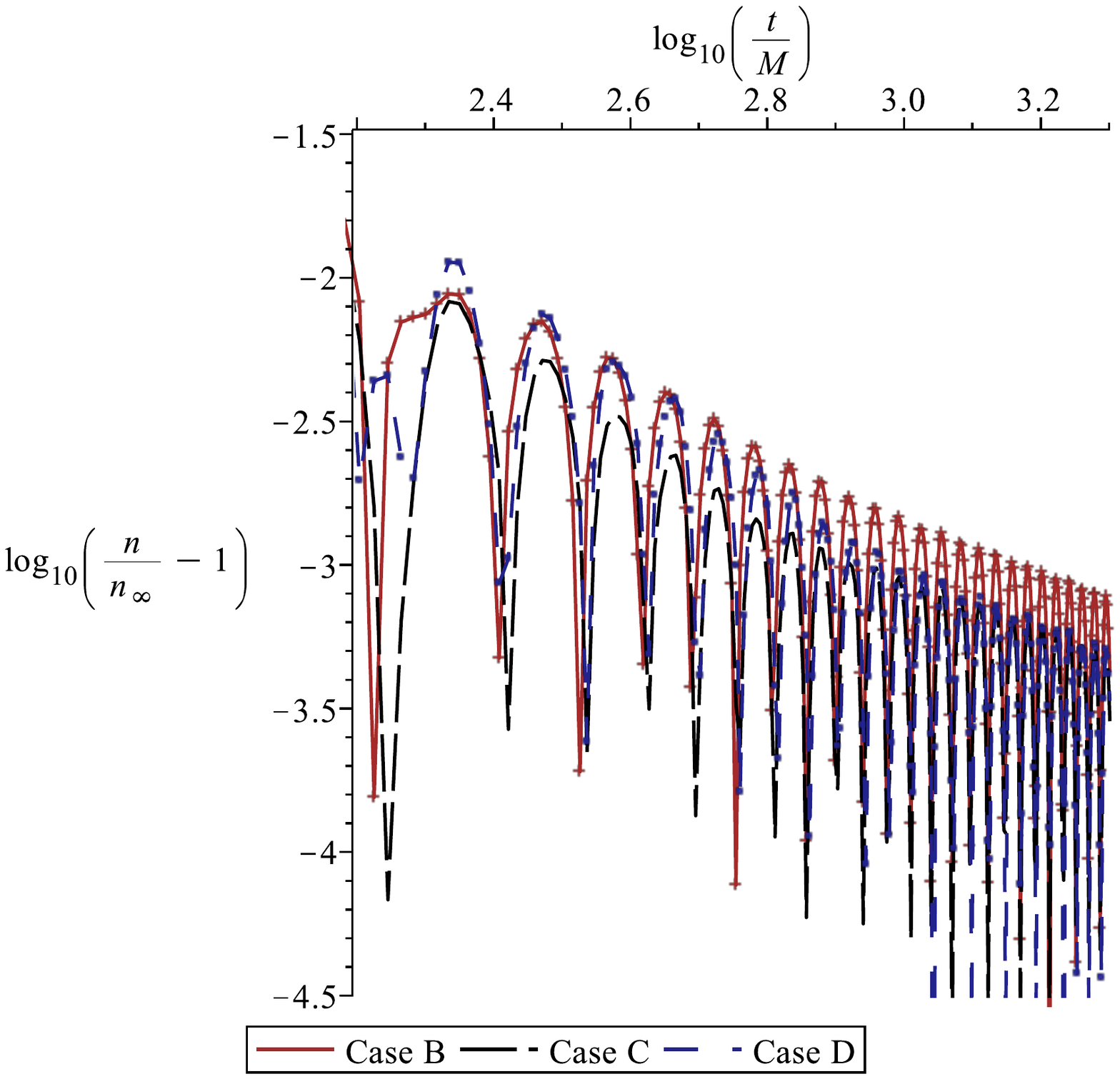}}
\resizebox{9.0cm}{!}{\includegraphics[trim= 36 160 36 60, clip=true]{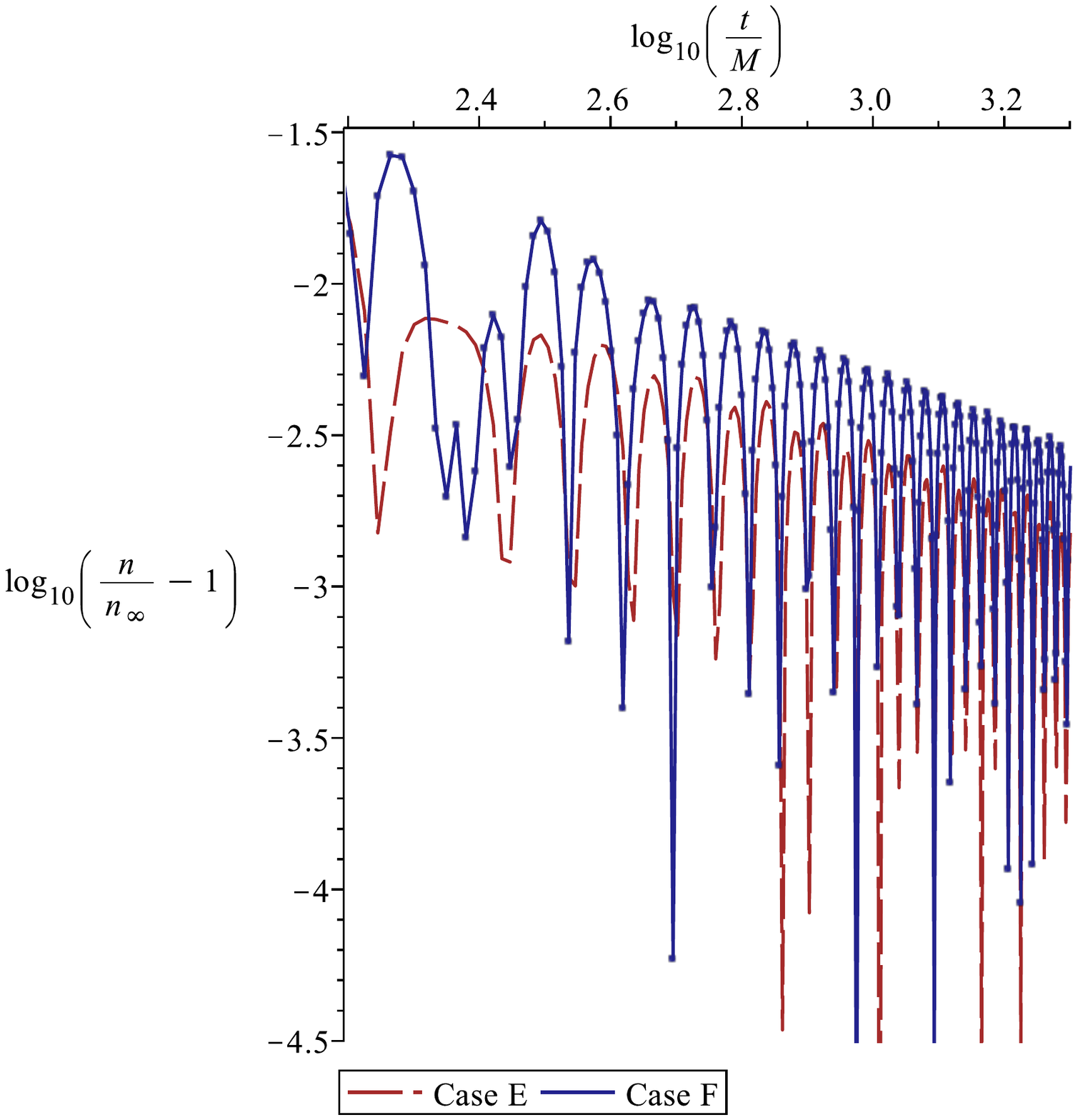}}
}
\caption{
$\log-\log$ plots of the relative error $|n/n_\infty - 1|$ for cases $B-F$, showing the decay of the particle density towards its final value $n_\infty$.}
\label{Fig:nC}
\end{figure}

In Figs.~\ref{Fig:nA}, \ref{Fig:nB} and~\ref{Fig:nC} we display the time behavior of the observer's particle density $n := {\cal J}^0_{\lambda > 0}$ for the case of an initial distribution function of the form $F(q^1,q^2,J_0,J_1,J_2) = F_Q(q^1,q^2) F_J(P,e)$, where
\begin{equation}
F_Q(q^1,q^2) := \frac{1}{N}
\exp\left[ -\frac{\cos^2(q^1)}{\eta_1^2} - \frac{\cos^2(q^2)}{\eta_2^2} \right],\qquad
F_J(P,e) := \exp\left[ -\frac{(P - 2e - P_0 + 2e_0)^2}{(\Delta P)^2} - \frac{(e - e_0)^2}{(\Delta e)^2} \right],
\label{Eq:FQFJ}
\end{equation}
$N$ is a positive normalization factor chosen such that $\int_0^{2\pi} dq^1 \int_0^{2\pi} dq^2 F_Q(q^1,q^2) = 1$, and $\eta_1,\eta_2,P_0,e_0,\Delta P,\Delta e$ are positive constants whose values are given in Table~\ref{Tab:Param}. Notice that the function $F_Q$ defined in Eq.~(\ref{Eq:FQFJ}) satisfies $F_Q(q^1,q^2) = F_Q(2\pi - q^1,2\pi - q^2)$, which implies that both terms in the sum $\sum_{\pm}$ yield the same contribution for ${\cal J}^\alpha_{\lambda > 0}$ when $\alpha=0,3$, while ${\cal J}^1_{\lambda > 0} = 0$. The observer is located at either $(r_{obs},\varphi_{obs}) = (6M,0)$ or $(9M,0)$.

\begin{table}
\begin{tabular}{|c|c|c|c|c|c|c|c|c|c|}
\hline
Case & $\eta_1$ & $\eta_2$ & $P_0$ &  $e_0$ & $(\Delta P)^2$ & $(\Delta e)^2$ & $\xi_{obs}$ & $\kappa$ & $t_{<0.001}$ \\
\hline
A & $3$ & $5$ & $9.2$  & $0.6$  & $0.4$ &  $0.1$ 
   & $6$ & - & $300M$ \\
B & $3$ & $5$ &$6.68$  & $0.14$ & $0.25$ &  $0.15$ 
   & $6$ & $0.9$ & $1500M$ \\
C & $3$ & $5$ &$6.3$  & $0.12$ & $0.25$ &  $0.2$ 
   & $6$ & $0.9$ & $900M$\\
D & $3$ & $5$ &$6.383$  & $0.092$ & $0.25$ &  $0.1$ 
   & $6$ & $0.9$ & $900M$ \\
E & $3$ & $5$ &$9.2$  & $0.1$ & $1.2$ &  $0.3$ 
   & $9$ & $0.8$ & $3500M$ \\
F & $3$ & $5$ &$9.2$  & $0.1$ & $0.4$ &  $0.1$ 
   & $9$ & $0.8$ & $7000M$ \\
\hline
\end{tabular}
\caption{Parameter values for the initial distribution function $F$ and the location of the observer $\xi_{obs}$. Also shown is the estimated power $\kappa$ obtained from fitting the envelope of the relative error $|n/n_\infty - 1|$ to an inverse power law of the form $t^{-\kappa}$, and the time $t_{<0.001}$ after which this error is estimated to lie below $0.001$. In case $A$ the decay is dominated by an initial exponential decaying phase, as indicated in the right panel of Fig.~\ref{Fig:nA}, after which the error is already very small, which makes it difficult to determine the power $\kappa$. In all other cases the values for $\kappa$ and $t_{<0.001}$ have been estimated to about $10\%$ accuracy.}
\label{Tab:Param}
\end{table}

Even though the distribution function oscillates in time and does not have a (pointwise) limit, the results in the plots indicate that the observable $n$ converges to a finite value (denoted by $n_\infty$) as $t\to \infty$. The theorem in the next section provides the means for determining this asymptotic value. For the moment we note that in case A the convergence is rather fast (exponential decay during the initial period, with relative fluctuations below $0.001$ after times larger than $300M$), while in the remaining cases the convergence appears to be slower (inverse power-law decay with relative fluctuations below $0.001$ after a few thousand $M$). We have also computed the remaining nontrival component ${\cal J}^3_{\lambda > 0}$ and the contributions ${\cal J}^\alpha_{\lambda < 0}$ belonging to negative angular momentum (assuming the same distribution $F_J(P,e)$) and verified that they exhibit a similar time behavior, with ${\cal J}^3 := {\cal J}^3_{\lambda > 0} + {\cal J}^3_{\lambda < 0}\to 0$.

\begin{figure}[ht]
\centerline{
\resizebox{6.0cm}{!}{\includegraphics[trim= 36 100 36 100, clip=true]{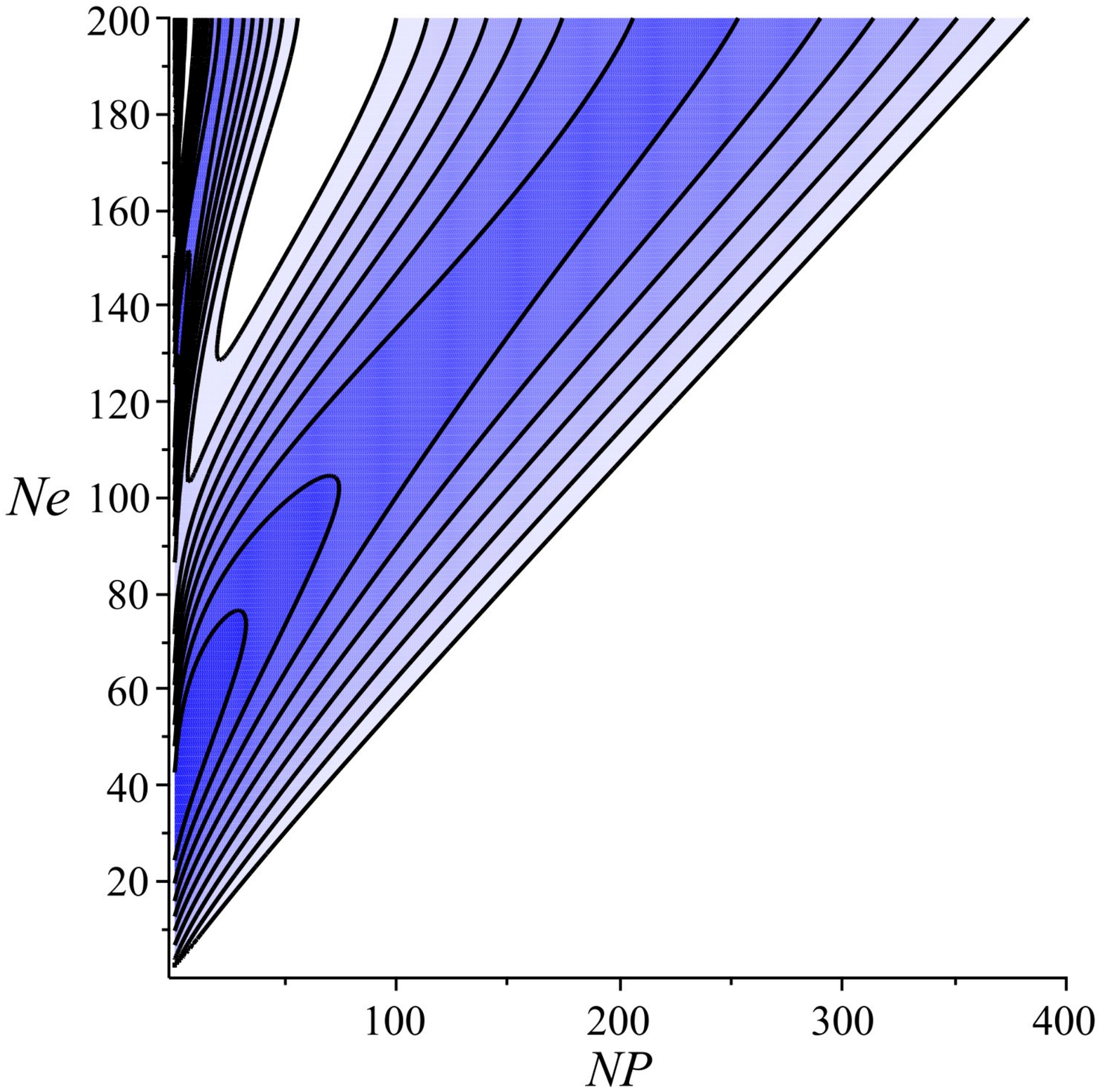}}
\resizebox{6.0cm}{!}{\includegraphics[trim= 36 100 36 100, clip=true]{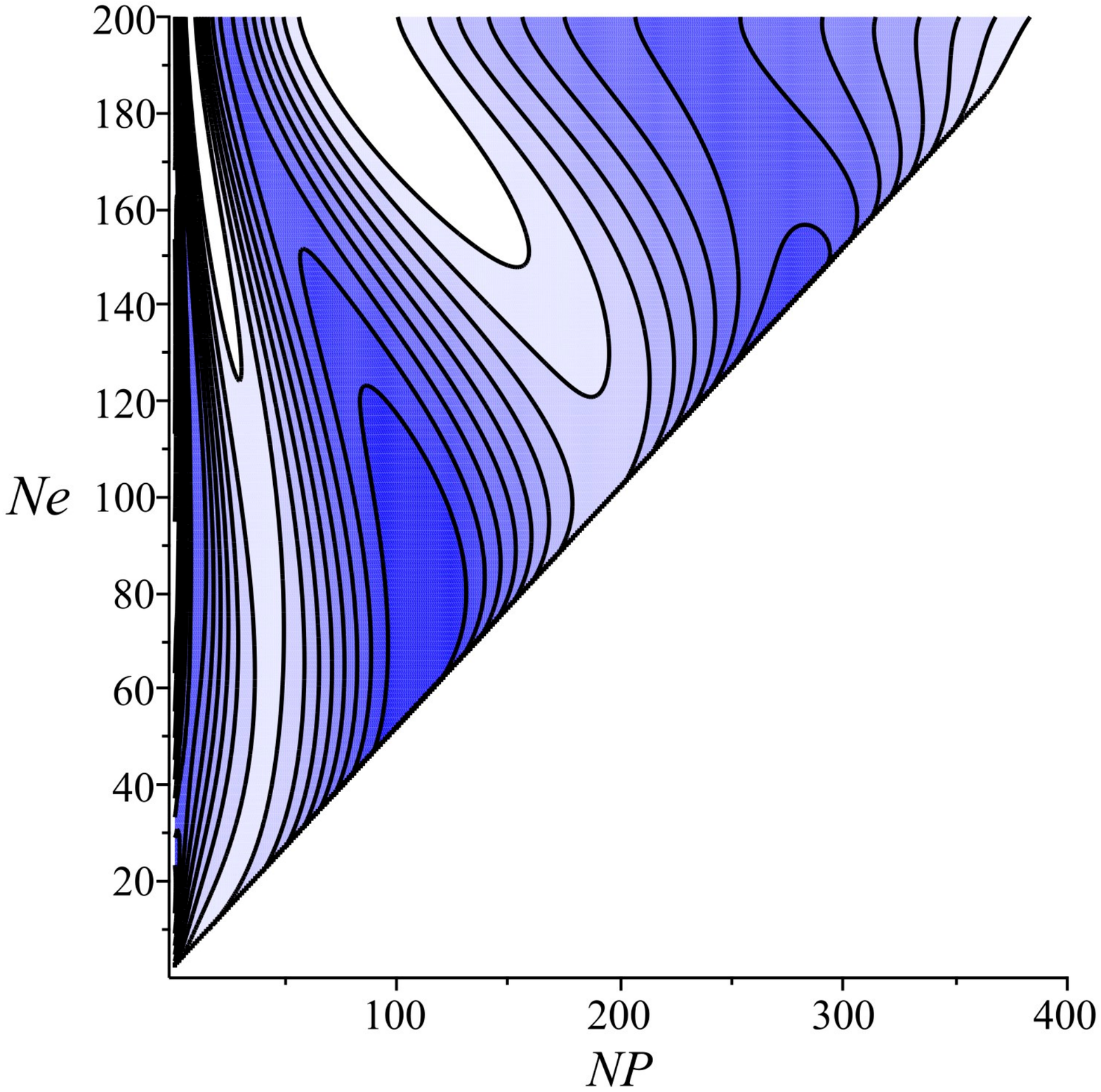}}
\resizebox{6.0cm}{!}{\includegraphics[trim= 36 100 36 100, clip=true]{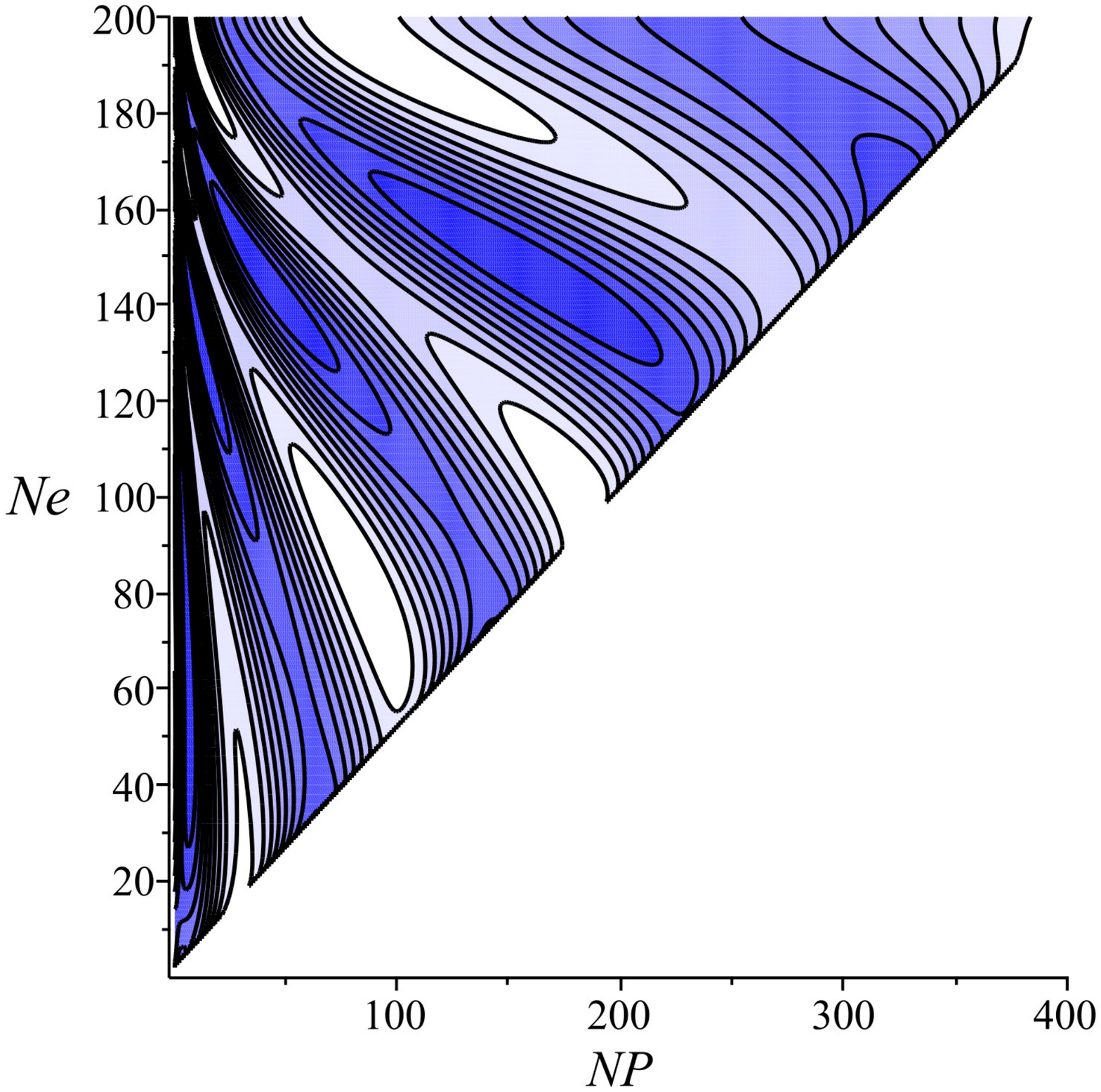}}
}
\centerline{
\resizebox{6.0cm}{!}{\includegraphics[trim= 36 100 36 100, clip=true]{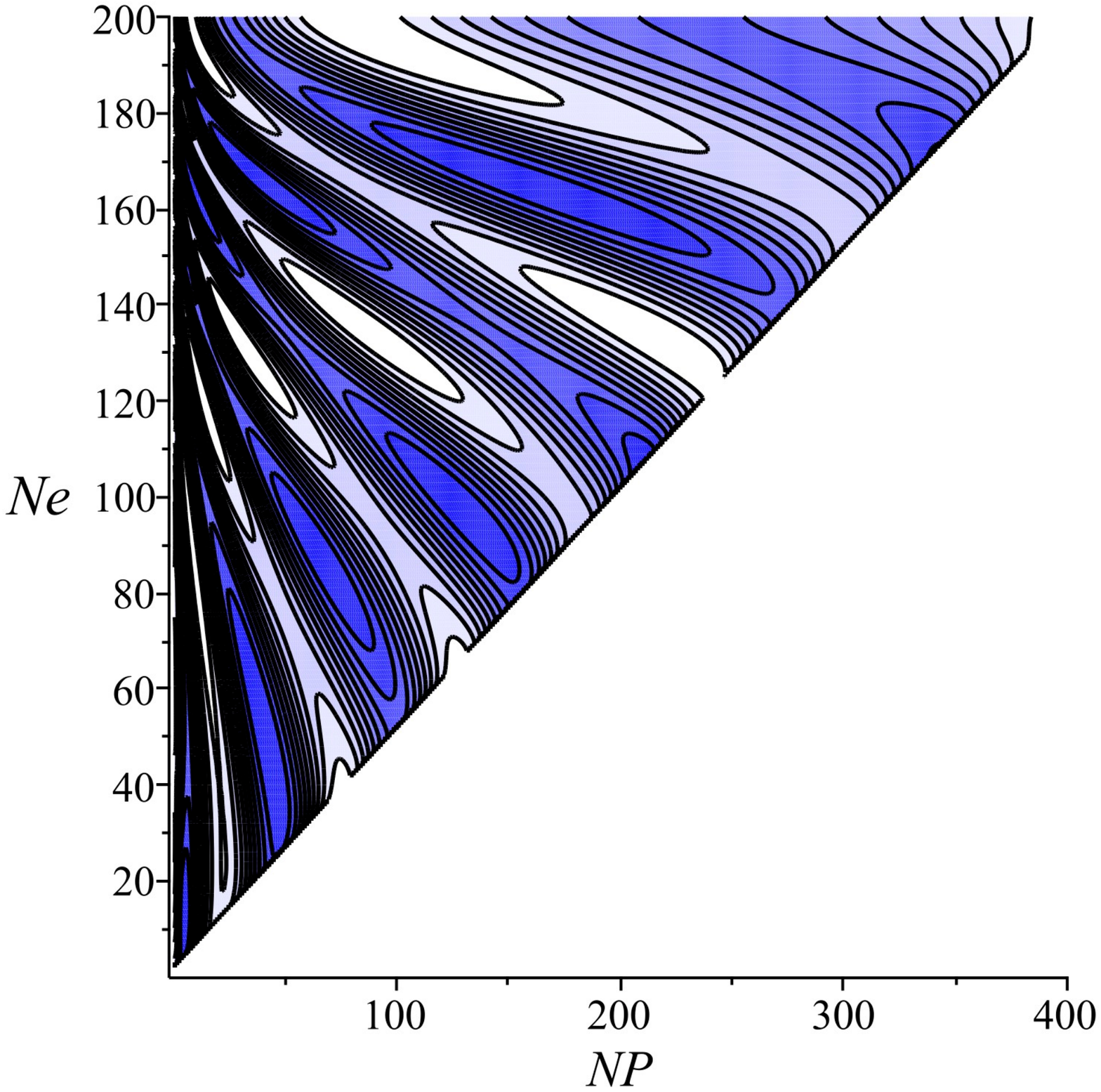}}
\resizebox{6.0cm}{!}{\includegraphics[trim= 36 100 36 100, clip=true]{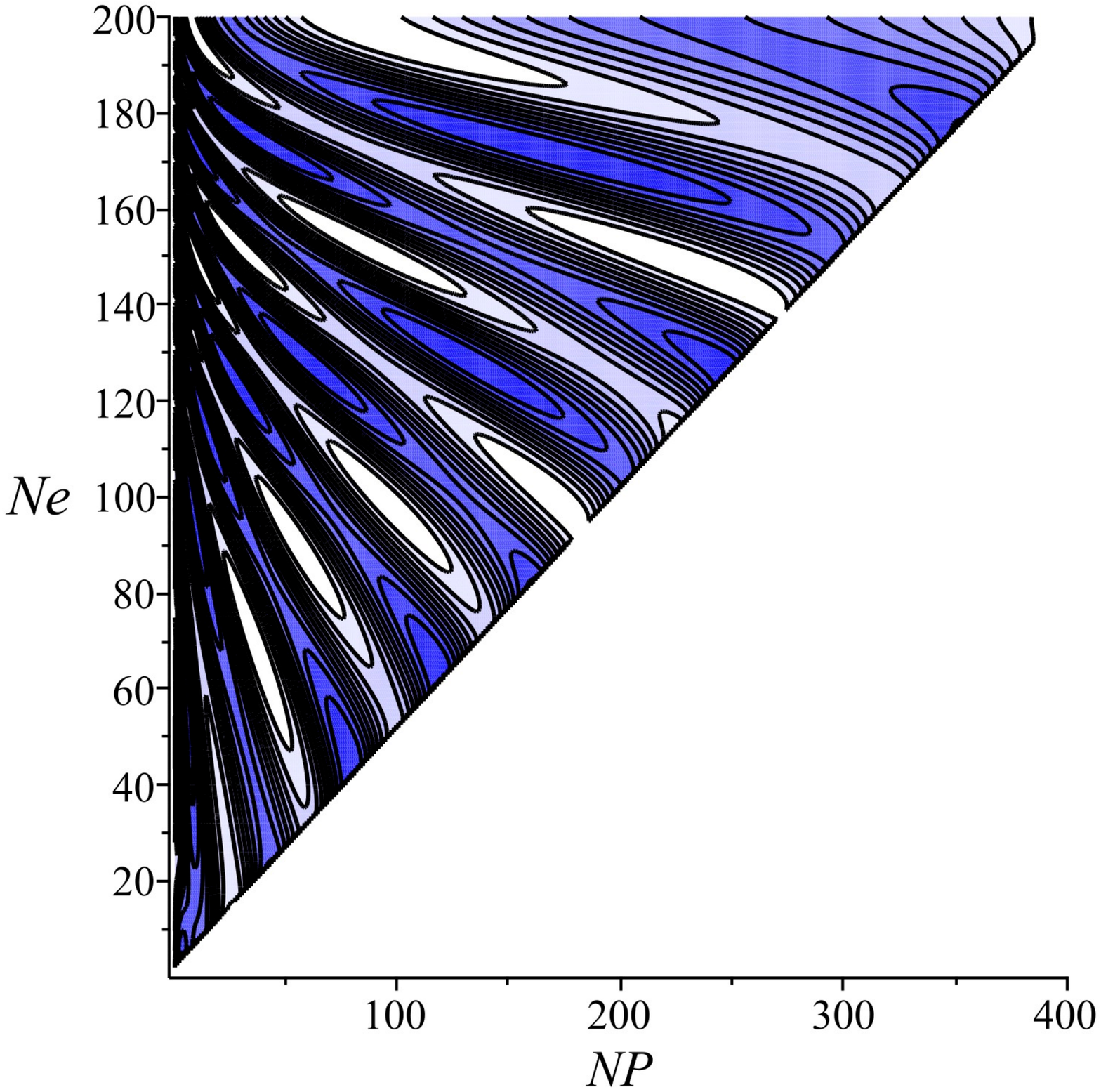}}
\resizebox{6.0cm}{!}{\includegraphics[trim= 36 100 36 100, clip=true]{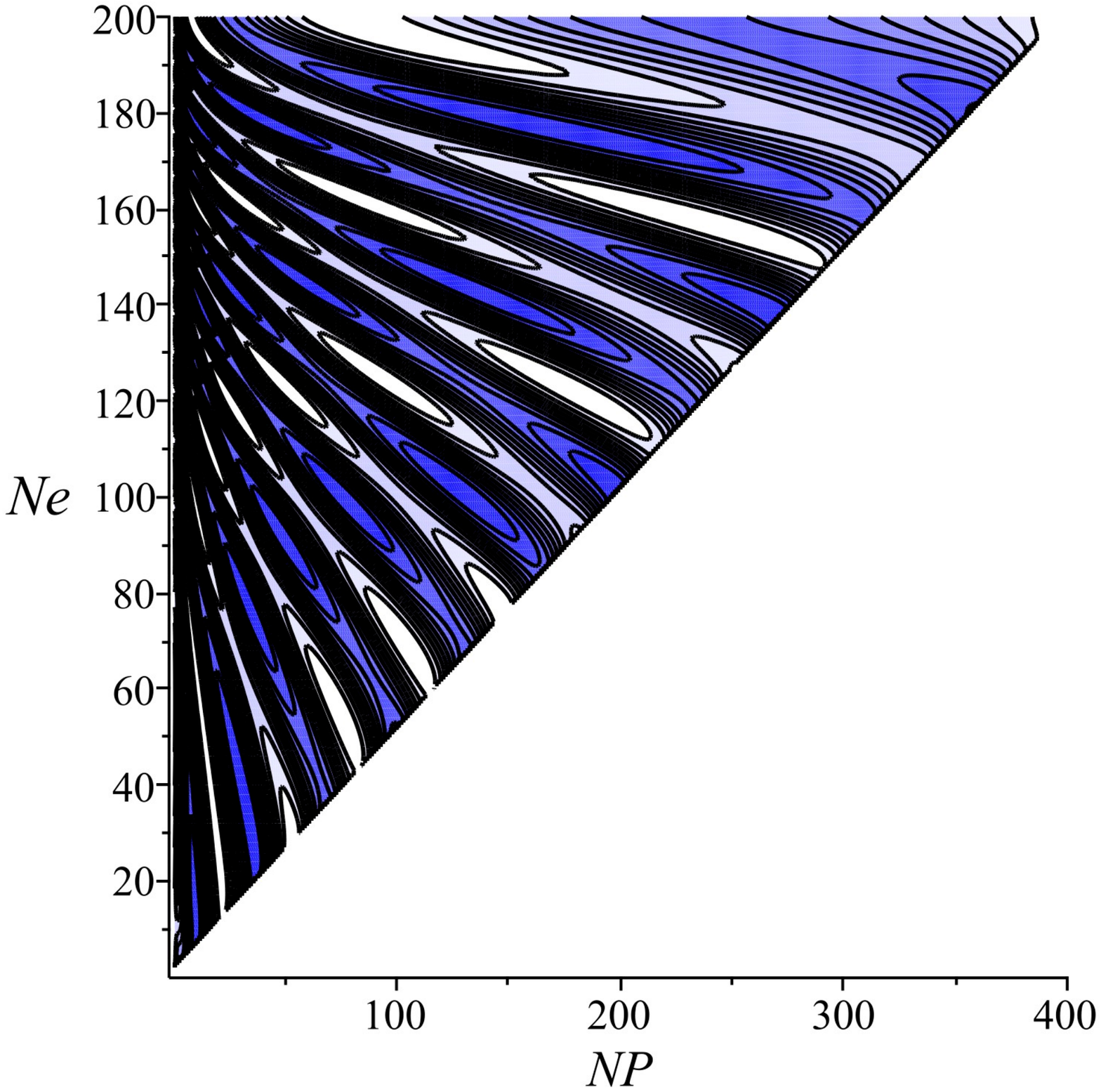}}
}
\caption{Level sets for the distribution function $F_Q(Q^1 -\omega^1 Q^0,Q^2 - \omega^2 Q^0)$ as a function of the parameters $(P':=P-6 - 2e,e)$ as seen by a static observer located at $r_{obs} = 6M$ and $\varphi_{obs} = 0$ for different times: $t = 0M$, $t = 100M$, $t = 200M$, $t = 300M$, $t = 400M$, and $t = 500M$ (from top left to right bottom). In these plots, the values of $NP$ and $Ne$ are related to $(P',e)$ through the formulas $P' = 0.01\times NP$ and $e = 0.005\times Ne$, respectively. The colors indicate different intervals in the range of $F_Q$, with dark blue corresponding to values of $F_Q$ lying close to its maximum and white corresponding to values of $F_Q$ close to its minimum.}
\label{Fig:PhaseSpace}
\end{figure}

As shown in the next section, the relaxation process displayed in Figs.~\ref{Fig:nA}-\ref{Fig:nC} is due to phase space mixing. To get an intuitive idea about this phenomenon, in Fig.~\ref{Fig:PhaseSpace} we show snapshots of the function $F_Q(Q^1 -\omega^1 Q^0,Q^2 - \omega^2 Q^0)$ for an observer located at $r_{obs} = 6M$ and $\varphi_{obs} = 0$ at different times. As is visible from these plots, the geodesic flow (which is volume preserving according to Liouville's theorem) stretches the phase space elements and spreads them over large regions in phase space, inducing the mixing property. As a consequence, averaged (macroscopic) quantities computed from the distribution function, such as the components of the current density in Eq.~(\ref{Eq:Jlambda}), have the form of an integral over a smooth function multiplied by an oscillating function whose frequency increases unboundedly in time. In the limit $t\to \infty$ these oscillations cancel out, and hence one can replace the distribution function with its non-oscillatory part, that is, its average over the angle variables:
\begin{equation}
F(Q^1 - \omega^1 Q^0, Q^2 - \omega^2 Q^0, J_0,J_1,J_2) \rightharpoonup 
\overline{F}(J_0,J_1,J_2) := \frac{1}{(2\pi)^2} \int\limits_0^{2\pi}\int\limits_0^{2\pi} F(Q^1,Q^2, J_0,J_1,J_2) dQ^1 dQ^2
\label{Eq:Convergence}
\end{equation}
The precise sense in which this convergence is valid will be explained in the next section. 


\section{Mathematical formulation of the mixing property}
\label{Sec:Proof}

After discussing the intuitive picture behind the mixing phenomenon and the corresponding relaxation process of the spacetime observables, in this section we provide a precise mathematical formulation of this effect which provides a rigorous explanation for the convergence of certain macroscopic observables. Here, a macroscopic observable is described by a test function $\varphi\in C_0^\infty(\Gamma_{bound})$ on the relativistic phase space $\Gamma_{bound}$, and its associated value is defined by the quantity
\begin{equation}
N[\varphi] := \int\limits_{\Gamma_{bound}} f(x,p) \varphi(x,p) \dvol_\Gamma,
\label{Eq:NDef}
\end{equation}
with $\dvol_\Gamma = dt d\varphi dr dp_t dp_\varphi dp_r$ the canonical volume form on $\Gamma_{bound}$ and $f(x,p)$ the one-particle distribution function describing the state of the kinetic gas. Note that our definition~(\ref{Eq:NDef}) is based on a fully covariant (i.e. independent of any choice of foliation or local coordinates) spacetime point of view which includes a time integral, such that $N[\varphi]$ does not  dependent on time. Therefore, in order to understand the dynamical behavior, one needs to perform a translation of the test function $\varphi$ along a time direction. Due to the many-fingered nature of time in general relativity there are no such preferred time directions in a general situation. In our case, the presence of the Killing vector field $\partial_t$ induces a natural vector field $\hat{k}$ on phase space $\Gamma_{bound}$ defined as the complete lift of $\partial_t$ (see Refs.~\cite{oStZ14b,pRoS16} for details). This vector field, in turn, induces a flow $\psi^t$ on $\Gamma_{bound}$ with respect to which one may define the time-translated test function:\footnote{Geometrically speaking, $\varphi_t$ is the push-forward of $\varphi$ with respect to $\psi^t$.}
\begin{equation}
\varphi_t(x,p) := \varphi(\psi^{-t}(x,p)).
\label{Eq:TimeTranslation}
\end{equation}
It is important to mention that in the rotating case, the Killing vector field $\partial_t$ is spacelike and not timelike inside the ergoregion, meaning that (due to the dragging by rotation) stationary observers lying inside the ergoregion cannot follow the integral curves of $\partial_t$. In this case one might replace $\partial_t$ with the linear combination $X := \partial_t + \Omega\partial_\varphi$, with the constant angular velocity $\Omega$ chosen such that $X$ is timelike in the vicinity of the observer, and define $\psi^t$ as the flow with respect to the complete lift $\hat{X}$ of $X$. This provides a more sensible definition for the time-translated test function $\varphi$, if $\varphi$ has its support in a region lying close to the event horizon which intersects the ergoregion. We stress that the theorem below holds for both (and probably more general) cases, the flow of $\psi^t$ being defined with respect to the complete lift of either $\partial_t$ or $X$.\footnote{As one can easily verify, the effect of replacing $\partial_t$ by $X$ is equivalent to replacing the frequency $\omega^1$ by $\omega^1 + \Omega$  in the proof of Theorem~\ref{Thm:Main}.}

For the statement of the following theorem, the area function, defined by 
\begin{equation} 
A_m(E,L) := \oint p_r dr = 2M m\left[ (1-\varepsilon^2)\HH_0 + \varepsilon \HH_3 \right],
\qquad L^2 > L_{ms}^2,\quad E_{min}(L) < E < E_{max}(L),
\label{Eq:AreaFunction}
\end{equation}
plays an important role. Note that this function determines the action variable $J_2 = \frac{A_m}{2\pi}$, and the frequencies $\omega^1$ and $\omega^2$ defined in Eq.~(\ref{Eq:FundamentalFrequencies}) are determined by the gradient of $A_m$ as follows:
\begin{equation}
\omega^1 =  \left. \frac{\partial A_m}{\partial L}  \middle/ \frac{\partial A_m}{\partial E} \right.,
\qquad
\omega^2 = -\left. 2\pi \middle/ \frac{\partial A_m}{\partial E} \right. .
\end{equation}
After these remarks, we are ready to formulate the main result of this article:

\begin{theorem}
\label{Thm:Main}
Let $F\in L^1(S^1\times S^1\times (-\infty,\infty)\times (-\infty,\infty)\times (0,\infty))$ be a Lebesgue-integrable function which, according to Eq.~(\ref{Eq:GeneralSolution}), determines a solution $f(x,p)$ on $\Gamma_{bound}$ of the collisionless Boltzmann equation on the equatorial plane of a Kerr black hole background. Let $\overline{F}$ and $\overline{f}$ be the corresponding distribution functions obtained by averaging over the angle variables. Let $\varphi\in C_0^\infty(\Gamma_{bound})$ be a test function, and denote by $\varphi_t$ its time-translation as defined in Eq.~(\ref{Eq:TimeTranslation}).

Suppose further that on the support of $\varphi$ the following non-degeneracy condition holds:
\begin{equation}
\det (D^2A_m(E,L)) \neq 0,
\label{Eq:NonDegeneracy}
\end{equation}
where $D^2 A_m(E,L)$ denotes the Hessian matrix of the area function~(\ref{Eq:AreaFunction}).

Then,
\begin{equation}
\lim\limits_{t\to \infty} N[\varphi_t] = \int\limits_{\Gamma_{bound}} \overline{f}(x,p) \varphi(x,p)\dvol_\Gamma.
\end{equation}
\end{theorem}

\begin{figure}[ht]
\centerline{
\resizebox{6.0cm}{!}{\includegraphics{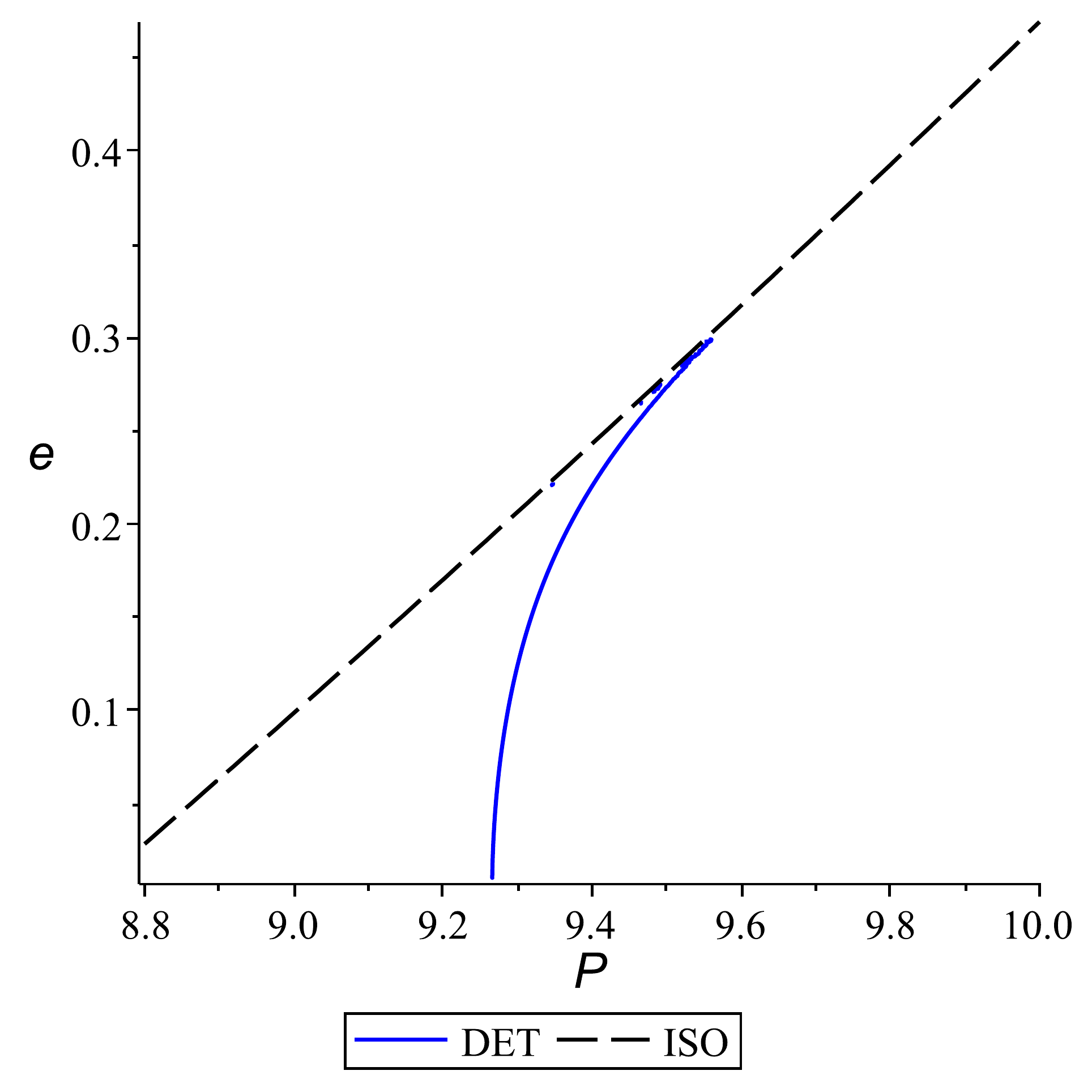}}
\resizebox{6.0cm}{!}{\includegraphics{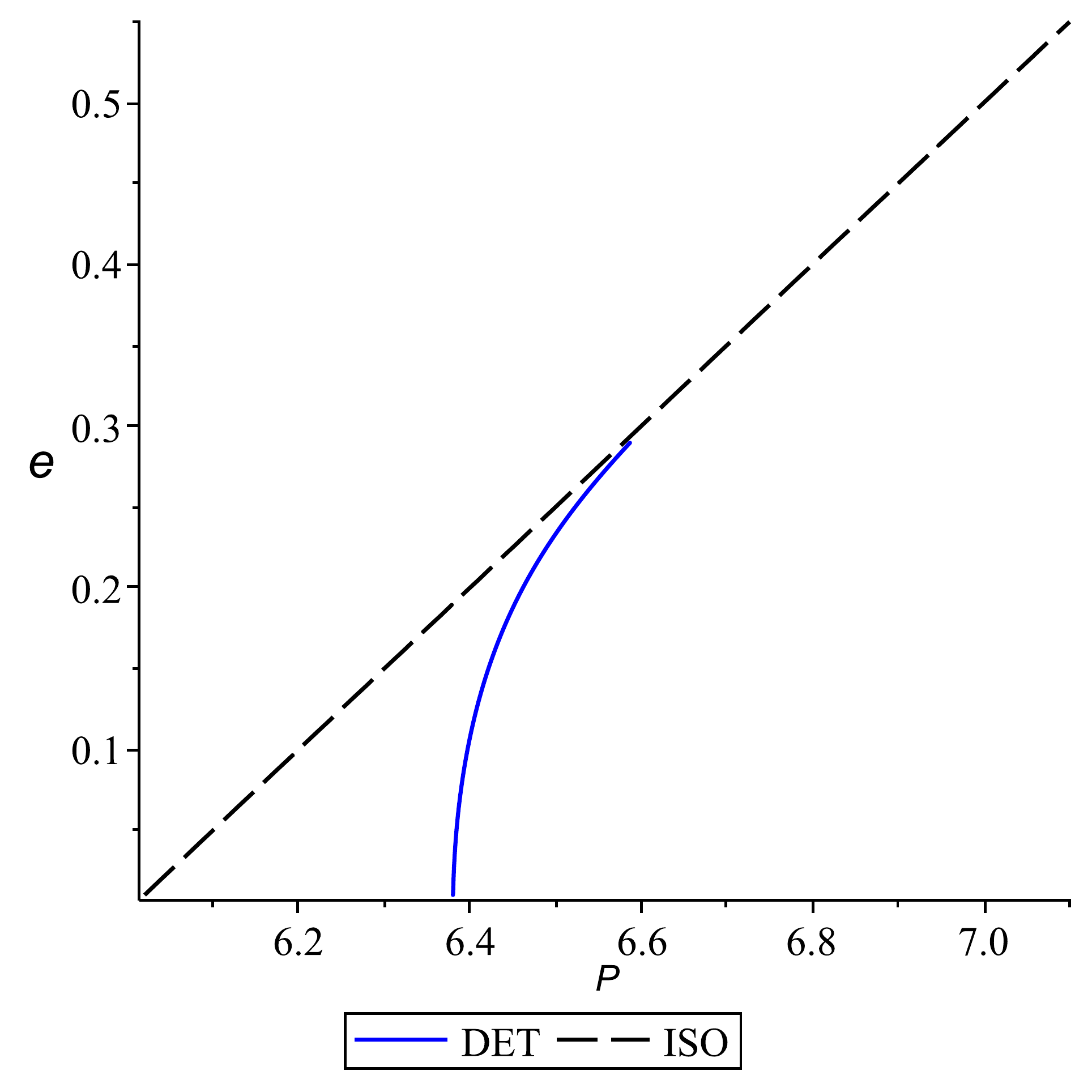}}
\resizebox{6.0cm}{!}{\includegraphics{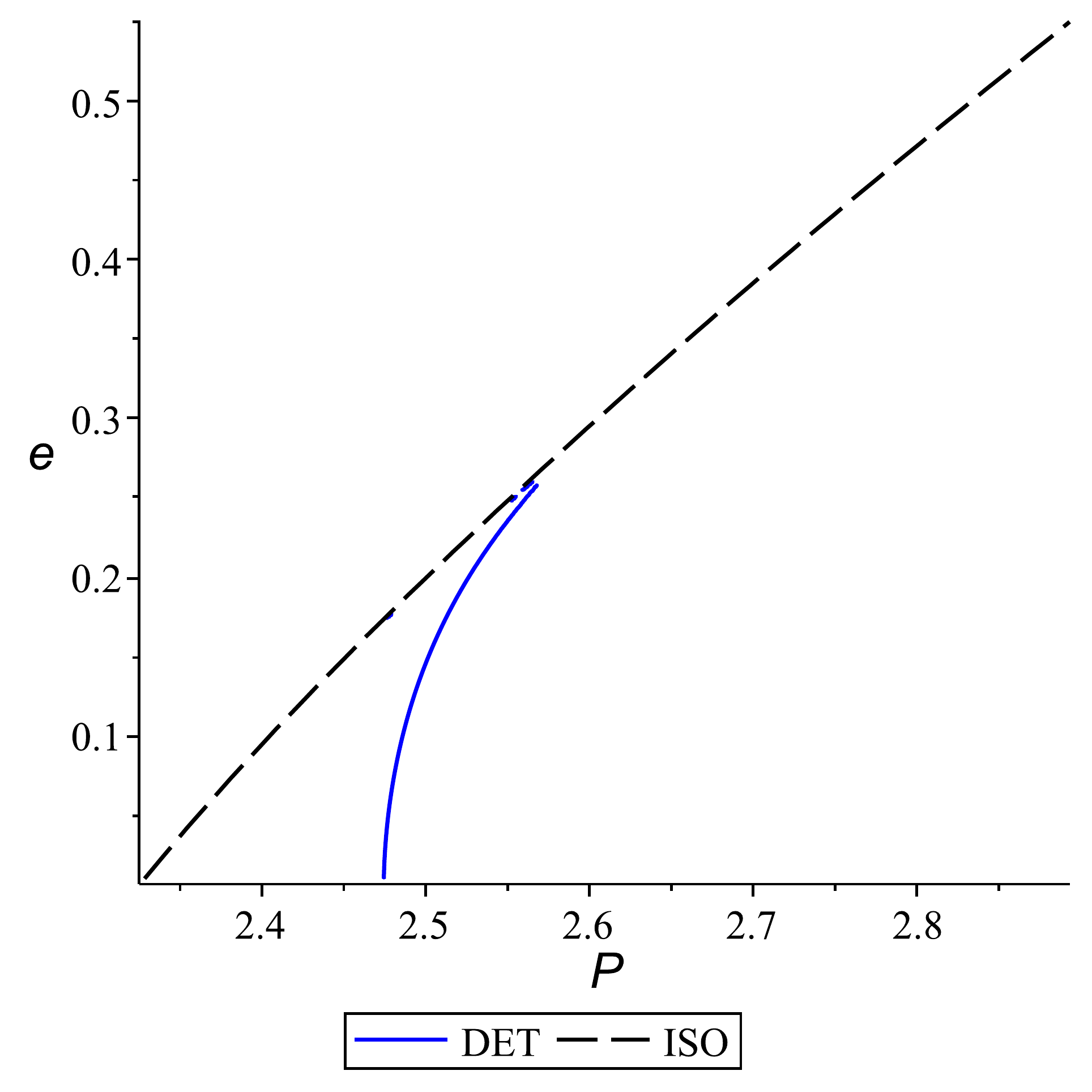}}
}
\caption{The determinant condition in the $(P,e)$-space for a Schwarzschild black hole (middle panel) and a Kerr black hole with rotational parameter $a = 0.9M$ (prograde orbits in the left panel, retrograde orbits in the right panel). The black dashed line corresponds to innermost stable orbits (ISO), for which $\xi_0 = \xi_1$, while the blue solid line indicates the points for which the determinant condition~(\ref{Eq:NonDegeneracy}) is violated.}
\label{Fig:DetCondition}
\end{figure}

{\bf Remarks}:
\begin{enumerate}
\item The validity of the determinant condition~(\ref{Eq:NonDegeneracy}) will be analyzed towards the end of this section, following the proof of the theorem. As will be verified numerically, it is satisfied everywhere except for points lying on a certain curve in $(E,L)$-space, see Fig.~\ref{Fig:DetCondition}. However, we stress that the theorem does not require condition~(\ref{Eq:NonDegeneracy}) to hold everywhere; it is sufficient to hold for points lying in the support of $\varphi$, that is, in the vicinity where the observer performs the measurement.

\item Although the above formulation of the theorem requires the test function $\varphi$ to be smooth, this assumption can be relaxed considerably. For example, it is possible to consider certain observables  with test functions of the form
$$
\varphi(x,p) = \delta(x - x_{obs})\sqrt{-\det(g^{\mu\nu}(x))} \psi(p),
$$
corresponding to spacetime observables at some given event $x_{obs}$, as the ones considered in the previous section, see~\cite{pRoS18b,pRoS18c}.

\item The quantity $n_\infty$ in Figs.~\ref{Fig:nA}, \ref{Fig:nB} and~\ref{Fig:nC} has been calculated by replacing $F$ by its average $\overline{F}$ in Eq.~(\ref{Eq:Jlambda}). 

\item Note that the hypothesis of the theorem require $\varphi$ to be supported on $\Gamma_{bound}$, which excludes circular orbits since the latter belong to the boundary of $\Gamma_{bound}$. However, the theorem can likely be generalized to the more realistic case in which the support of $\varphi$ includes circular orbits (see also the remarks at the end of this section).

\end{enumerate}

\proofof{Theorem~\ref{Thm:Main}} In a first step, we rewrite the integral defining $N[\varphi]$ in terms of the action-angle variables $(Q^\alpha,J_\alpha)$ as follows:
$$
N[\varphi] = \int F(Q^1 - \omega^1 Q^0, Q^2 - \omega^2 Q^0, J_0,J_1,J_2) 
\Phi(Q^0,Q^1,Q^2,J_0,J_1,J_2) dQ^0 dQ^1 dQ^2 dJ_0 dJ_1 dJ_2,
$$
where $\Phi(Q^0,Q^1,Q^2,J_0,J_1,J_2) = \varphi(x,p)$ is the representation of the test function $\varphi$ in terms of $(Q^\alpha,J_\alpha)$ and where we have used the fact that the transformation $(x,p)\mapsto (Q^\alpha,J_\alpha)$ is canonical. With respect to these variables, the flow associated with the vector field $\hat{k} = -\partial/\partial Q^0$ amounts to a translation of $Q^0$ by $-t$, keeping the other variables fixed, such that $\varphi_t(x,p) = \Phi(Q^0 + t,Q^1,Q^2,J_0,J_1,J_2)$. By means of the variable substitution $Q^0 = \Theta^0 - t$, $Q^A = \Theta^A + \omega^A\Theta^0$, $A = 1,2$, one obtains
$$
N[\varphi_t] = \int\limits_0^\infty dm \int\limits_0^{2\pi} d\Theta^1 \int\limits_0^{2\pi} d\Theta^2 
\int\limits_{L^2 > L_{ms}^2} dL \int\limits_{E_{min}(L)}^{E_{max}(L)} dE\, 
F_m(\Theta^1 + \omega^1 t, \Theta^2 + \omega^2 t, E,L) 
\Phi_m(\Theta^1,\Theta^2, E,L),
$$
where we have defined $F_m(\Theta^1,\Theta^2,E,L) := F(\Theta^1,\Theta^2,E,L,A_m(E,L)/2\pi)$ and
$$
\Phi_m(\Theta^1,\Theta^2,E,L) 
 := \frac{1}{2\pi}\frac{\partial A_m}{\partial m}(E,L)
\int\limits_{-\infty}^\infty \Phi\left( \Theta^0,\Theta^1 + \omega^1\Theta^0,\Theta^2 + \omega^2\Theta^0,E,L,\frac{A_m(E,L)}{2\pi} \right)  d\Theta^0.
$$
In a next step, we perform a variable substitution in order to replace the integral over $(E,L)$ by an integral over  $(\omega^1,\omega^2)$. To this purpose, we denote by $W_m: (E,L)\mapsto (\omega^1,\omega^2)$ the map that defines (for fixed $m$) the angular frequencies $(\omega^1,\omega^2)$ in terms of the constants of motion $(E,L)$. Its Jacobian determinant is related to the determinant of the Hessian of the area function as follows:
$$
\det(D W_m) = -\frac{(\omega^2)^3}{(2\pi)^2} \det (D^2A_m(E,L)).
$$
Therefore, due to condition~(\ref{Eq:NonDegeneracy}) the map $W_m$ is (at least) locally invertible on the support of $\varphi$. If not globally invertible, we may cover its domain with open subsets $U_i$ on which $W_m$ is invertible. Since $\Phi$ is compactly supported, only a finite number of these subsets are required. On each of these $U_i$'s we define
$$
G_{m,i}(\Theta^1,\Theta^2,\omega^1,\omega^2) := \left\{ \begin{array}{ll}
\frac{F_m(\Theta^1,\Theta^2,E,L))}{|\det DW_m(E,L)|} , & (\omega^1,\omega^2)\in W_m(U_i),\\
0, & \hbox{otherwise}.
\end{array} \right.
$$
Assuming first that $\Phi_m$ is supported in one of these sets $U_i$, we have
\begin{equation}
N[\varphi_t] =  \int\limits_0^\infty dm \int\limits_0^{2\pi} d\Theta^1 \int\limits_0^{2\pi} d\Theta^2 
\int\limits_{\Real^2} d\omega^1 d\omega^2 g_{m,i}(t,\Theta^1,\Theta^2,\omega^1,\omega^2)
 \Phi_m(\Theta^1,\Theta^2,E,L),
\label{Eq:NphitAA}
\end{equation}
where
\begin{equation}
g_{m,i}(t,\Theta^1,\Theta^2,\omega^1,\omega^2) := G_{m,i}(\Theta^1 + \omega^1 t,\Theta^2 + \omega^2 t,\omega^1,\omega^2).
\end{equation}

After these initial steps, we encounter ourselves exactly in the same situation as a collisionless gas in a periodic box, and the mixing property is easily revealed by means of Fourier transformation (see section 3 in~\cite{cMcV11}): for this, define
$$
\hat{G}_{m,i}(k_1,k_2,\eta_1, \eta_2) := \frac{1}{(2\pi)^2} \int\limits_{-\infty}^\infty  \int\limits_{-\infty}^\infty \int\limits_0^{2\pi} \int\limits_0^{2\pi}G_{m,i}(\Theta ^1 , \Theta ^2,\omega ^1, \omega ^2)
 e^{-i(k_1 \Theta ^1 + k_2 \Theta ^2)} e^{- i(\eta_1\omega^1 + \eta_2\omega^2)}  
  d\Theta ^1 d \Theta ^2 d\omega ^1 d\omega ^2 ,
$$
where $(k_1, k_2 ) \in \Integer^2$ and $(\eta_1, \eta_2) \in\Real^2$. Then,
$$
\hat{g}_{m,i}(t,k_1, k_2,\eta_1, \eta _2) = \hat{G}_{m,i}(k_1,k_2,\eta_1 - k_1 t, \eta _2 - k_2 t),
$$ 
hence the Fourier transform converts the rotations of the angle variables $(\Theta^1,\Theta^2)\mapsto (\Theta^1 + \omega^1 t,\Theta^2 + \omega^2 t)$ into a translation $(\eta_1,\eta_2)\mapsto (\eta_1 - k_1 t,\eta_2 - k_2 t)$ of the frequencies associated with the angular frequencies. According to the Riemann-Lebesgue lemma $\hat{g}_{m,i}(t,k_1, k_2,\eta _1, \eta _2)$ converges pointwise to $0$ for all fixed $(k_1, k_2,\eta _1, \eta _2)$ with $(k_1,k_2) \neq (0,0)$. Consequently,
$$
\lim\limits_{t\to \infty} \hat{g}_{m,i}(t,k_1, k_2, \eta _1, \eta _2) = \delta_{k_1 0} \delta_{k_2 0}\hat{G}_{m,i}(0,0,\eta_1, \eta _2) =: \hat{g}_{m,i}^{\infty}(k_1, k_2,\eta _1, \eta_2),
$$
for all $(k_1, k_2,\eta _1, \eta _2)\in \Integer^2\times \Real^2$ with the convergence rate determined by the smoothness of $G_{m,i}(\Theta ^1, \Theta ^2,\omega ^1, \omega ^2)$ with respect to $\omega ^{1}, \omega ^2$. Here, $\hat{g}_{m,i}^{\infty}(k_1, k_2,\eta _1, \eta_2)$ is the Fourier transform of the function
$$
g^{\infty}_{m,i}(\Theta ^1, \Theta ^2,\omega ^1, \omega^2) = \frac{1}{(2\pi)^2}\int\limits_0^{2\pi} \int\limits_0^{2\pi} G_{m,i}(\Theta ^1, \Theta ^2,\omega^1, \omega^2) d\Theta ^1 d\Theta^2 
 = \overline{G_{m,i}}(\Theta ^1, \Theta ^2,\omega^1, \omega^2),
$$
which is the average of $G_{m,i}$ over the angle variables. Since $\Phi_m$ is smooth and has compact support, it follows that $N[\varphi_t]$ converges for $t\to \infty$ to the same expression as the one in the right-hand side of in Eq.~(\ref{Eq:NphitAA}) with $g_{m,i}(t,\Theta^1,\Theta^2,\omega^1,\omega^2)$ replaced with $g^{\infty}_{m,i}(\Theta ^1, \Theta ^2,\omega ^1, \omega^2)$. This proves the theorem for the case in which the support of $\Phi_m$ lies within one of the subsets $U_i$.

For the general case, a partition of unity can be used to write $\Phi_m$ as a finite sum of functions each of which is supported in only one of the subsets $U_i$.
\qed

Now that the theorem has been proven, we discuss the validity of the non-degeneracy condition~(\ref{Eq:NonDegeneracy}). Based on the explicit representation in Eq.~(\ref{Eq:AreaFunction}), we have computed the determinant of the Hessian of the area function. As in the previous section, it is convenient to describe the result in terms of the generalized Keplerian variables $(P,e)$ instead of $(E,L)$. Interestingly, it turns out the determinant condition is satisfied everywhere except for points lying on the blue solid curves shown in Figs.~\ref{Fig:DetCondition} for different values of the rotational parameter.

To provide analytic support for the results shown in these figures we consider the particular case of quasi-circular orbits on a Schwarzschild background, for which $\alpha = 0$ and $e\ll 1$. In this case, the expansion of the first derivatives of the area function in terms of the eccentricity $e$ yield
\begin{eqnarray}
A_E &:=& \frac{\partial A_m}{\partial E} = M\frac{2\pi P^2}{\sqrt{P-6}}
\left[ 1 + \frac{3}{4}\frac{2P^3 - 32P^2 + 165P - 266}{(P-2)(P-6)^2} e^2 + {\cal O}(e^4) \right],
\label{Eq:AE}\\
A_L &:=& \frac{\partial A_m}{\partial L} = -\frac{2\pi\sqrt{P}}{\sqrt{P-6}} 
\left[ 1 + \frac{3}{4}\frac{1}{(P-6)^2} e^2 + {\cal O}(e^4) \right],
\label{Eq:AL}
\end{eqnarray}
giving
$$
\det\left( \begin{array}{cc} \frac{\partial A_E}{\partial P} &   \frac{\partial A_E}{\partial e} \\
 \frac{\partial A_L}{\partial P} &   \frac{\partial A_L}{\partial e} \end{array} \right)
 = M D(P) e + {\cal O}(e^3),\qquad
 D(P) := -9\pi^2 \frac{P^{3/2} (4P^2 - 39P + 86)}{(P-2)(P-6)^3},\qquad P > 6.
$$
The function $D$ is positive when $P$ is slightly larger than $6$ and negative for large $P$, and it has a single root at $P = P_* := (39 + \sqrt{145})/8 \simeq 6.38$, which corresponds to the limit point of the blue solid curve as $e\to 0$ in the middle panel of Fig.~\ref{Fig:DetCondition}. From Eqs.~(\ref{Eq:AE},\ref{Eq:AL}) one can also conclude that $(A_E,A_L)$ is a function of $(P,e^2)$ which is locally invertible for small $e^2$ and $P$ away from $P_*$.

We end this section by observing that the distribution functions considered in the previous section in cases $A$, $E$ and $F$ have their main support away from the blue solid curve in $(P,e)$-space where the determinant condition is violated, while in the remaining cases $B$, $C$ and $D$ the main support of the distribution function intersects this curve (compare the values given in Table~\ref{Tab:Param} with the middle panel of Fig.~\ref{Fig:DetCondition}). Although the determinant condition is violated in the latter cases, the plots in Figs.~\ref{Fig:nB} and \ref{Fig:nC} suggest that the particle density still converges for $t\to \infty$, indicating that mixing is still sufficiently strong for the relaxation process to take place.

\section{Conclusions}
\label{Sec:Conclusions}

In this work, we have shown that a relativistic, collisionless kinetic gas propagating in the equatorial plane of a Kerr black hole spacetime settles down to a stationary, axisymmetric configuration. As we have demonstrated, this relaxation process is due to phase space mixing, an effect that plays a prominent role in a wide range of fields in physics including galactic dynamics and plasma physics. Here, we have exhibited the relevance of the mixing phenomenon for the dynamical behavior of spacetime observables within the fully general relativistic setting of a kinetic gas which is trapped in the strong field gravitational field of a black hole.

The main implication of this work is that the one-particle distribution function $f$ describing the state of the gas, which in general is a function of the five coordinates $(Q^1,Q^2,J_0,J_1,J_2)$ parametrizing the relativistic phase space $\Gamma_{bound}$ corresponding to bound trajectories, can be replaced by a much simpler distribution function $\overline{f}$ which is a function depending only of the constants of motion $(J_0,J_1,J_2)$ and which can be computed by averaging the initial distribution function over the angle variables $(Q^1,Q^2)$. Indeed, our main theorem in the previous section shows that, provided the determinant condition~(\ref{Eq:NonDegeneracy}) is satisfied on the support of $\varphi$, the integral over $f$ times any test function $\varphi$ converges in time to the integral over $\overline{f}$ times the same test function $\varphi$. At the physical level, these integrals have the interpretation of macroscopic observables, where different choices for the test function correspond to different physical quantities being measured.

The determinant condition~(\ref{Eq:NonDegeneracy}) means that the map $(E,L)\mapsto (\omega^1,\omega^2)$ which defines the angular frequencies in terms of the constants of motion $E$ and $L$ is locally invertible, a condition that is well-known in perturbation theory of integrable Hamiltonian systems, see for example Section X.51 in~\cite{Arnold-Book}. For geodesic motion in the equatorial plane of a Kerr black hole we have shown that the determinant condition holds everywhere with the exception of points in the $(E,L)$-plane lying on a curve (a zero-measure set) which connects a particular circular orbit to innermost stable orbits at large eccentricities. As the plots in Section~\ref{Sec:Relaxation} indicate, mixing still occurs in the vicinity of these exceptional points, suggesting that our theorem also holds under weaker assumptions. It should be interesting to relate the behavior of the gas in the vicinity of the exceptional points to Tremaine's analysis of stable singularities or catastrophes in galaxies~\cite{sT99} and the implications on the time scale in which the mixing occurs. In any case, for the specific examples we have analyzed in Section~\ref{Sec:Relaxation}, the damping of the oscillations in the particle density is rather fast, with relative amplitude lying below $0.001$ after a few thousand light-crossing times corresponding to the gravitational radius of the black hole.

A further interesting problem consists in analyzing the effects of the self-gravity of the gas configuration (which have been neglected in the present work) on the mixing property. The inclusion of the self-gravity implies that the Kerr metric acquires correction terms due to the non-vanishing stress-energy tensor associated with the gas, which in turn leads to a perturbed Hamiltonian flow describing the geodesic motion for the gas particles. Based on general arguments from Kolmogorov-Arnold-Moser (KAM) theory, it has been argued~\cite{jBmGtH13,jBmGtH15} that such perturbations could trigger dynamical chaos in the vicinity of resonant orbits. Therefore, it should be particularly interesting to study the mixing phenomenon in the neighborhood of such orbits, and investigate whether or not the relaxation of the observables persists in the self-gravitating case. 

The generalization of the mixing property to thick disk configurations, where individual gas particles are not necessarily confined to the equatorial plane, will be given in future work~\cite{pRoS18c}.


\acknowledgments

We wish to thank Francisco Astorga, Michael Kiessling, Alexander Komech, Ulises Nucamendi, and Thomas Zannias for fruitful and stimulating discussions. We also thank Michael Kiessling, and Thomas Zannias for comments on a previous version of this manuscript. This research was supported in part by CONACyT Grants No. 577742, by the CONACyT Network Project No. 294625 ``Agujeros Negros y Ondas Gravitatorias", and by a CIC Grant to Universidad Michoacana. We also thank the Erwin Schr\"odinger Institute for Theoretical Physics, where this work was initiated, for hospitality. P.R. also thanks Chalmers University of Technology and the University of Vienna for hospitality, where part of this work was completed.

\appendix
\section{Explicit expressions for the functions $\HH_j$}
\label{App:HHFunctions}

Using the definition of the function $R(r)$ in Eq.~(\ref{Eq:EnergySurface}), the following relations between $(\varepsilon,\lambda)$ and its roots $(\xi_j)$ are obtained:
\begin{equation}
1 - \varepsilon^2 = \frac{2}{\xi_{012}},\qquad 
\hat{\lambda}^2 = \frac{\xi_0 \xi_1 \xi_2}{\xi_{012}},\qquad
\hat{\lambda}^2 + 2\alpha\hat{\lambda}\varepsilon + \alpha^2 = 2\frac{\xi_0 \xi_1 + \xi_0 \xi_2 + \xi_1 \xi_2}{\xi_{012}},
\label{relations_rootsE}
\end{equation}
where we have abbreviated $\xi_{012} := \xi_0 + \xi_1 + \xi_2$ and introduced $\hat{\lambda} := \lambda - \alpha\varepsilon$. The functions  $\HH_j(\chi)$ and corresponding constants $\HH_j := \HH_j(\pi/2)$ in terms of which the angle variables $Q^\alpha$ are expressed are defined as follows:
\begin{eqnarray}
\HH_0(\chi) &:=& -\frac{C}{2}\Bigg\{Ê(\xi_0 \xi_{012} - \xi_1 \xi_2)\FF(\chi,k)
 + \xi_1(\xi_2 - \xi_0)\EE(\chi,k) \Bigg.\nonumber\\
&& \left. + (\xi_1 - \xi_0) \xi_{012} \Pi(\chi,b^2,k) - \frac{1}{2} \xi_1(\xi_2 - \xi_1)
\frac{\sqrt{1 - k^2\sin^2\chi}}{1 - b^2\sin^2\chi}\sin(2\chi) \right\},
\\
\HH_1(\chi) &:=& -C\left\{
\left( \hat{\lambda} + \alpha\frac{\varepsilon \xi_0^2 - \alpha\hat{\lambda}}{(\xi_0 - \xi_+)(\xi_0 - \xi_-)} \right)\FF(\chi,k)
\right.\nonumber\\
&& \left. - \alpha\frac{\xi_1 - \xi_0}{\xi_+ - \xi_-}\left[ \frac{\varepsilon \xi_+^2 - \alpha\hat{\lambda}}{(\xi_0 - \xi_+)(\xi_1 - \xi_+)} \Pi(\chi,b_+^2,k) - (+\leftrightarrow -) \right] \right\},
\\
\HH_2(\chi) &:=& 2C\Bigg\{ \frac{\xi_0(\varepsilon \xi_0^2 - \alpha\hat{\lambda})}{(\xi_0 - \xi_+)(\xi_0 - \xi_-)}
\FF(\chi,k) + \varepsilon(\xi_1 - \xi_0)\Pi(\chi,b^2,k) \Bigg.\nonumber\\
&& \left. - \frac{\xi_1 - \xi_0}{\xi_+ - \xi_-}\left[ \frac{\xi_+(\varepsilon \xi_+^2 - \alpha\hat{\lambda})}{(\xi_0 - \xi_+)(\xi_1 - \xi_+)}\Pi(\chi,b_+^2,k) - (+\leftrightarrow -) \right] \right\},
\end{eqnarray}
and
\begin{eqnarray}
\HH_3(\chi) &:=& \HH_2(\chi) + \frac{\lambda}{\varepsilon}\HH_1(\chi)
\nonumber\\
 &=& \frac{2C}{\sqrt{\xi_{012}(\xi_{012}-2)}}\Bigg\{
 \xi_0^2\FF(\chi,k) + (\xi_1 - \xi_0)(\xi_{012} - 2)\Pi(\chi,b^2,k) \Bigg.\nonumber\\
 && \left. - \frac{\xi_1 - \xi_0}{\xi_+ - \xi_-}
 \left[ \xi_+(\xi_2 - \xi_+)\Pi(\chi,b_+^2,k) - (+\leftrightarrow -) \right] \right\},
\end{eqnarray}
with $b := \sqrt{(\xi_2 - \xi_1)/(\xi_2 - \xi_0)}$, $k := \sqrt{\xi_0/\xi_1} b$, $\xi_\pm := 1 \pm \sqrt{1 - \alpha^2}$, $b_\pm := \sqrt{\frac{\xi_0 - \xi_\pm}{\xi_1 - \xi_\pm}} b$ and $C := \sqrt{\frac{2\xi_{012}}{\xi_1(\xi_2-\xi_0)}}$, and where the functions $\FF(\chi,k)$, $\EE(\chi,k)$ and $\Pi(\chi,b^2,k)$ denote standard elliptic integrals as defined, for instance, in Ref.~\cite{DLMF}.

In terms of these functions one finds
\begin{equation}
{\cal S} = M m\left[ -\varepsilon\frac{t}{M} + \lambda \varphi +  (1-\varepsilon^2)\HH_0(\chi) + \varepsilon \HH_3(\chi) \right],
\end{equation}
and
\begin{equation}
\frac{\partial {\cal S}}{\partial m} = M\HH_0(\chi),\qquad
\frac{\partial {\cal S}}{\partial L} = \varphi + \HH_1(\chi),\qquad
\frac{\partial {\cal S}}{\partial E} = -t + M\HH_2(\chi) - E\HH_0(\chi),
\end{equation}
from which the expressions~(\ref{Eq:QDef}) are easily derived.

\bibliographystyle{unsrt}

\end{document}